\documentclass[12pt]{article}
\pdfoutput=1
\usepackage{putex}
\usepackage{amsmath}
\usepackage{amssymb}

\usepackage{graphicx}
\usepackage[lofdepth,lotdepth]{subfig}
\usepackage{epstopdf}
\usepackage{enumerate}
\usepackage{cite}
\usepackage{tensor}
\usepackage{slashed}
\usepackage{feynmf}
\usepackage{hyperref}
\usepackage{bbold}
\usepackage{dsfont}
\usepackage{mathrsfs}
\usepackage{caption}
\usepackage{comment}

\usepackage{youngtab}
\usepackage{amsfonts}
\usepackage{braket}

\renewcommand{\theequation}{\thesection.\arabic{equation}}

\usepackage[utf8x]{inputenc}
\usepackage{braket}
\usepackage{titlesec}
\usepackage{fixltx2e}
\usepackage{cleveref}
\hypersetup{breaklinks}

\usepackage[usenames,dvipsnames,svgnames,table]{xcolor}

\def\p{\partial}

\newcommand {\be} {\begin {equation}}
\newcommand {\ee} {\end {equation}}
\newcommand{\bea}{\begin{eqnarray}}
\newcommand{\eea}{\end{eqnarray}}
\newcommand{\dDisc}{\text{dDisc}}
\newcommand{\F}{{}_2F_1}
\newcommand\numberthis{\addtocounter{equation}{1}\tag{\theequation}}

\def\zb{\overline{z}}

\def\rt{\rightarrow}

 \newmuskip\pFqmuskip

\newcommand*\pFq[6][8]{%
  \begingroup 
  \pFqmuskip=#1mu\relax
  \mathcode`\,=\string"8000
  \begingroup\lccode`\~=`\,
  \lowercase{\endgroup\let~}\pFqcomma
  {}_{#2}F_{#3}{\left[\genfrac..{0pt}{}{#4}{#5};#6\right]}%
  \endgroup
}
\newcommand{\pFqcomma}{\mskip\pFqmuskip}

\makeatletter
\renewcommand{\@maketitle}{
\newpage
 \begin{center}%
  {\large\bfseries \@title \par}%
 \end{center}%
 \par} \makeatother

\numberwithin{equation}{section}

\titleformat*{\section}{\large\bfseries}

\begin{document}

\institution{UCLA}{ \quad\quad\quad\quad\quad\quad\quad\quad\quad Mani L. Bhaumik Institute for Theoretical Physics, \cr Department of Physics and Astronomy, University of California, Los Angeles, CA 90095, USA}

\title{Light-state Dominance from the Conformal Bootstrap}

\authors{Per Kraus${ }^\beta$ \let\thefootnote\relax\footnote{\texttt{${}^\beta$pkraus@physics.ucla.edu}} and Allic Sivaramakrishnan${ }^z$ \let\thefootnote\relax\footnote{\texttt{${}^z$allic@physics.ucla.edu}}}

\abstract{
We derive forms of light-state dominance for correlators in CFT$_d$, making precise the sense in which correlators can be approximated by the contribution of light operator exchanges. Our main result is that the four-point function of operators with dimension $\Delta$ is approximated, with bounded error,  by the contribution of operators with scaling dimension below $\Delta_c > 2\Delta$ in the appropriate OPE channel. Adapting an existing modular invariance argument, we use crossing symmetry to show that the heavy-state contribution is suppressed by a relative factor of $e^{2\Delta-\Delta_c}$. We extend this result to the first sheet and derivatives of the correlator. Further exploiting technical similarities between crossing and modular invariance, we prove analogous results for the $2d$ partition function along the way. 

We then turn to effective field theory in gapped theories and AdS/CFT, and make some general comments about the effect of integrating out heavy particles in the bulk.    Combining our bounds with the Lorentzian OPE inversion formula we show that, under certain conditions, light-state dominance implies that integrating out heavy exchanges leads to higher-derivative couplings suppressed at large $\Delta_{gap}$.

}

\date{}

\maketitle
\setcounter{tocdepth}{2}
\tableofcontents

\section{Introduction}

A cornerstone of our understanding of quantum field theory is the idea that low energy physics is insensitive to short distance details, or more precisely that the dependence on such details can be absorbed into the values of a finite number of low energy coupling constants. This idea is made concrete and useful in general settings by effective field theory and the renormalization group. Conformal field theories (CFTs) are scale invariant, and so possess no absolute notion of low energy versus high energy. However, individual observables have a characteristic energy scale. For a given observable, such as a correlation function of local operators, it is meaningful to ask about the sensitivity to details of the CFT data at high energy, i.e. the dependence on the spectrum and OPE coefficients of high dimension operators. This question is the topic of the present work.

In the case of large $N$ CFTs, the AdS/CFT correspondence connects the standard notion of low energy effective field theory in the bulk, where the energy scale is measured relative to the string or Planck scale, to corresponding notions in the CFT.  In particular, such holographic CFTs have some type of gap in their operator spectrum, and one can ask about sensitivity to details of the CFT data above the gap.  In a global sense, crossing symmetry implies that the low and high dimension CFT data are not independent, but the question considered here is the dependence on high dimension data of a fixed observable, such as a correlation function of low dimension operators at fixed locations.  We henceforth refer to high and low dimension operators as being ``heavy" and ``light" respectively. We also use the term ``light-state dominance" to refer to situations in which quantities are determined by the light CFT data, up to small corrections.

A simple and familiar context for light-state dominance is the thermodynamics of 2d CFTs, as captured by the partition function $Z(\beta)$.   Modular invariance, $Z(\beta) = Z({4\pi^2 \over \beta})$, relates the high and low energy spectra to one another.  For example, the asymptotic density of states --- the Cardy formula \cite{CardyFormula} --- is fixed in terms of the ground state energy.   This allows one to estimate the contribution of states above some suitably large cutoff dimension $\Delta_c$.  Modular invariance can similarly be applied to extract results on the asymptotic spectral density weighted by OPE coefficients \cite{KrausM16,ChangL16,CardyMM17,KellerMZ17,ChoCY17,DasDP17,Romero-BermudezSS18,HikidaKT18,BrehmDD18,Kusuki118,Kusuki218}, as well as for CFTs defined on spaces of amenable topologies in higher dimensions \cite{Shaghoulian15,Shaghoulian16}.

Crossing symmetry similarly controls the asymptotically heavy contributions to four-point functions $G(z, \bar{z})$ on the plane. In \cite{PappadopuloRER12, RychkovY15}, crossing and tauberian theorems were used to estimate the rate of convergence of the OPE, and this allows one to estimate the contribution of heavy states, $\Delta> \Delta_c$, for asymptotically large $\Delta_c$.  This approach was extended using the complex tauberian theorem in \cite{MukhametzhanovZ18}, in which corrections to the heavy-state estimates were computed. It was empirically found that these estimates were in good agreement with low-$\Delta_c$ contributions in certain known correlators, despite being derived only for asymptotically large $\Delta_c$. In \cite{KimKO15} a different approach was taken to derive bounds for the relative contribution of the heavy states. Here, upper bounds were obtained for $\Delta_c \rightarrow \infty $ in various limiting cases of $\Delta, c, d$ where $c$ is the $2d$ central charge and $d$ is the spacetime dimension. Bounds were also derived for the correlator at $z = \bar{z} = 1/2$ for finite $\Delta_c, \Delta$ satisfying $2<4\Delta<\Delta_c$. In \cite{LinSSWY15}, the correlator at $z = \bar{z} = 1/2$ was bounded above by $\frac{2^{1+2\Delta} \Delta}{\Delta_{min}-2\Delta} $, where $\Delta_{min}$ is the dimension of the lightest exchanged operator, assumed to obey $\Delta_{min}>2\Delta$. Other analytic bootstrap methods provide information about various families of operators \cite{FitzpatrickKPS12,KomargodskiZ12,Simmons-Duffin16,LargeSpinPerturbationTheory,OPEInversion}, but do not produce constraints on the total contribution of all heavy operators.

Hartman, Keller, and Stoica (HKS) took a different approach and gained access to the contribution of operators above some finite (as opposed to asymptotic) cutoff, $\Delta_c > \epsilon+c/12$ with $\epsilon >0$. Modular invariance was used to obtain inequalities involving the full light (L) and heavy (H) contributions to the partition function $Z(\beta) = Z_L(\beta) + Z_H(\beta)$. HKS showed that
\begin{equation}
\log Z(\beta) =
\left\{
\begin{array}{ll}
\log Z_L(\beta) + \mathcal{E}(\beta), & \beta >2\pi, \\
\log Z_L(\beta') + \mathcal{E}(\beta'),& \beta < 2\pi,
\end{array}  \right.
\label{HKS2.11}
\end{equation}
where $\beta' = 4\pi^2/\beta$, and the contribution from heavy states is bounded by $\mathcal{E}(\beta) \leq - \ln \left(1-  e^{(\beta'-\beta)\epsilon} \right) $.  This is especially useful at large central charge, since $\log Z_L  \sim c$, while the error terms are order $c^0$.  By further assuming a sparse light spectrum, HKS showed that the CFT thermodynamics is governed by the extended Cardy regime expected in holographic theories. In particular, the leading part of the  CFT free energy agrees with that of thermal AdS and BTZ solutions for $\beta > 2 \pi$ and $ \beta < 2\pi$ respectively. The HKS procedure was used to extend these results to thermal correlators in \cite{KrausSS17}, in which a modular crossing condition was used to show that one and two-point functions are well-approximated by the contributions from operators with $ \Delta < \epsilon+c/12$. Under the appropriate light-state sparseness condition, this proved that generalized free operators ($\Delta$ fixed as $c \rightarrow \infty$) are well-approximated by the contribution from only the generalized free sector, and that these correlators agree with those computed in thermal AdS and BTZ backgrounds for $\beta > 2 \pi$ and $ \beta < 2\pi$ respectively. The HKS argument has been further applied to the case $\beta_L \neq \beta_R$ in \cite{AnousMS18}. These results demonstrate that universal features of gravity in AdS$_3$ at large $c$ follow from light-state dominance   in CFT$_2$ under the appropriate assumptions about light state data, corroborating general expectations that light-state sparseness is a necessary CFT condition for an Einstein gravity dual.

We will study light-state dominance for four-point correlators in $d$-dimensional unitary CFTs.  Applying the OPE yields an expression for the correlator in terms of a sum over exchanged operators. The scenario to be ruled out is one in which the number of heavy operators or the size of their OPE coefficients is large  enough to prevent any good light state approximation to the correlator. For example, there might be an OPE coefficient of a single heavy operator that is large enough to contribute appreciably.  In that case, we might need to know the spectrum and OPE coefficients of arbitrarily heavy operators in order to compute $G(z,\zb)$ to reasonable accuracy. Our results rule out this possibility by showing it is inconsistent with crossing symmetry.

We make progress by adapting the HKS approach to the four-point function of identical scalars with dimension $\Delta$, obtaining a bound on the heavy state contribution that is valid at finite, not asymptotically large, $\Delta_c$. Surprisingly, the elementary but powerful HKS technique goes through for the correlator with little modification. We prove various forms of light-state dominance in $d$-dimensional unitary CFTs for all $\Delta_c > 2\Delta$ without making any assumptions about the theory. Setting $z = \bar{z}$ for convenience, the essential result for the correlator is simple and universal:
\begin{equation}
G(z) =
\left\{
\begin{array}{ll}
 ~~~~~~~~~~~G_L(z) \left(1+\mathcal{E}(z) \right) ,& z < 1/2, \\
\left(\frac{z}{1-z} \right)^{2\Delta}  G_L(1-z) \left(1+\mathcal{E}(1-z)\right) , & z > 1/2,
\end{array}  \right.
\end{equation}
where the contribution from heavy states is bounded according to $ \mathcal{E}(z)  \leq \frac{R(z)}{1-R(z)}$ and $R(z) = \left(\frac{z}{1-z} \right)^{\Delta_c-2\Delta} $. This bound applies to $z\in [0,1]$, and a stronger bound becomes relevant at $z = 1/2$. More generally, we obtain bounds everywhere on the first $(z,\zb)$ sheets as well as even derivatives of the correlator. Along the way, we derive analogous results for $Z(\beta)$.

One theme in this work is that modular invariance and crossing symmetry share  technical similarities that can often be exploited to perform analogous computations for the partition function and correlator. The similarities can provide an intuitive way to understand our results, and so we summarize several here.
\begin{align*}
e^{-\beta} &~~~\sim~~~ z
\\
E_{vac} &~~~\sim~~~ -2\Delta
\\
\beta = 2\pi &~~~\sim~~~ z = 1/2
\\
\beta \rightarrow \infty &~~~\sim~~~ z\rightarrow 0
\\
\beta \rightarrow 0 &~~~\sim~~~ z\rightarrow 1
\\
e^{-\beta} \rightarrow e^{-4\pi^2/\beta} &~~~\sim~~~ z \rightarrow 1-z
\\
\text{Cardy formula \cite{CardyFormula} }  &~~~\sim~~~ \text{OPE convergence estimates \cite{PappadopuloRER12}}
\\
\text{Light-state dominance for $Z(\beta)$ \cite{HartmanKS14} }  &~~~\sim~~~ \text{This work}
\end{align*}
The differences between results for the correlator and partition function can be traced back to the difference between the modular and crossing transformations.

More generally, we hope to make clear that the HKS approach is not special to $Z(\beta)$ or even modular invariance, but can be applied to a wide variety of ``duality invariant" objects. The amenable quantities we study have two competing singularities in the domain of the continuous duality parameter and can be written as a sum over Boltzmann-type factors that transform non-trivially under the duality. We find that the HKS method is fairly robust, as the essential result holds in a variety of applications, and succeeds without certain ingredients of the original HKS argument, namely monotonicity of the Boltzmann-type factors and positivity of the terms in the light part of the sum.

It is useful to compare and contrast the idea of approximating a CFT correlator by dropping heavy states to that of integrating out a heavy particle in QFT.  In the latter case, at least as an asymptotic expansion in the inverse heavy mass, the effect of virtual heavy particles can be reproduced by contact interactions built out of the light fields.  The coefficients of the non-renormalizable interactions go to zero as the heavy mass is taken to infinity.  On the other hand, a heavy operator exchange in a CFT correlation function cannot in any sense be reproduced be readjusting the light data.  This is related to the fact that the contribution of a heavy state decays in an exponential, as opposed to power law, fashion as the heavy dimension is taken large.  This distinction is made clearer in the context of AdS/CFT in Witten diagrams for the exchange of a heavy particle in the bulk.  As the mass is taken large, there are power-suppressed contributions that can be reproduced by bulk contact interactions, and exponentially suppressed contributions that cannot.  The CFT counterpart is that such a correlator involves the exchange of both heavy and light operators.  The light (double trace) operator contribution is the power law suppressed part that can be matched to the effect of the bulk contact interactions.   Keeping these facts in mind is important to appreciate the distinction between placing bounds on the contributions of heavy CFT operators versus heavy bulk particles.

A well-known conjecture \cite{HeemskerkPPS09} is that a large $N$ CFT with a gap in its spectrum of single trace operators (typically defined by the twist of the lightest spin $j>2$ single trace operator) will have an AdS dual description that is local down to a scale set by $1/\Delta_{gap}$. For this to be the case, the bulk theory should be such that integrating out heavy particles above the gap induces contact interactions whose coefficients are in accord with effective field theory lore; namely, the coefficients are suppressed by inverse powers of $\Delta_{gap}$, with one additional power for each derivative.

This conjecture was addressed using the Lorentzian OPE inversion formula \cite{OPEInversion}.  The idea here is to relate the contact interactions to double trace OPE coefficients, which are in turn related by the inversion formula to the double discontinuity of the correlator.  The positivity and boundedness of the double discontinuity leads to a bound on the contribution of heavy operators to the OPE coefficients. The bound involves the spin of the contact interaction rather than its dimension, where the spin refers to the maximal spin of two-particle states that couple to the operator.   The distinction between heavy operators and heavy bulk particles is immaterial at bulk tree level, since the double discontinuity  projects out the light double trace operators that arise from an exchange Witten diagram.  There are a number of subtleties in this chain of reasoning, and we revisit the problem using our heavy state bound.    We work with a finite cutoff $\Delta_c$, chosen to coincide with $\Delta_{gap}$, and then obtain bounds on OPE coefficients that decays as $\Delta_{gap}$ is taken large. This approach is relatively direct and simple; however, it does not provide us with detailed information about the rate of decay with $\Delta_{gap}$ of the contact interaction coefficients.

The remainder of this paper is organized as follows. In section \ref{sec:background}, we review relevant background. We derive the central light-state dominance result  in section \ref{sec:main_result}. We extend this result to the first $(z, \bar{z})$ sheets in section \ref{sec:first_sheet} and to derivatives in section \ref{sec:derivatives}. We explore various applications of light-state dominance in section \ref{sec:applications}, addressing effective field theory and gapped theories with a focus on AdS/CFT.

\section{Background}
\label{sec:background}

We will consider the four-point function of identical scalar primaries $\mathcal{O}$ with dimension $\Delta$,
\begin{equation}
\braket{\mathcal{O}(x_1)\mathcal{O}(x_2)\mathcal{O}(x_3)\mathcal{O}(x_4)}
=\frac{G(u,v)}{(x_{12})^{2\Delta} (x_{34})^{2\Delta}},
\end{equation}
with cross ratios
\begin{equation}
u=\frac{x_{12}^2 x_{34}^2}{x_{13}^2 x_{24}^2}, ~~~~ v=\frac{x_{14}^2 x_{23}^2}{x_{13}^2 x_{24}^2}.
\end{equation}
Crossing symmetry follows from demanding invariance under interchanging $x_2\leftrightarrow x_4$,
\begin{equation}
u^{-\Delta} G(u,v) = v^{-\Delta} G (v,u).
\label{BootstrapEquation}
\end{equation}
It is often convenient to map the operator insertions to $0,(z,\bar{z}),1,\infty$, in which case $u = z \bar{z}, v = (1-z)(1-\bar{z})$. We will sometimes mix $(z, \bar{z})$ and $(u,v)$ notation for compactness and refer to $(u,v) \rightarrow (v,u)$ as the crossing transformation. $G$ has the expansion
\begin{equation}
G(z,\bar{z}) = \sum_{\Delta_p, l_p} C_{\mathcal{O} \mathcal{O} \mathcal{O}_p}^2 g_{\Delta_p,l_p}(z, \bar{z})
\end{equation}
in terms of conformal blocks $g_{\Delta_p,l_p}(z, \bar{z})$ and OPE coefficients $C_{\mathcal{O} \mathcal{O} \mathcal{O}_p}$. The OPE coefficients $C_{\mathcal{O} \mathcal{O} \mathcal{O}_p}$ vanish unless the spin $l_p$ is even. A basis can be chosen so $\langle \mathcal{O}_p(1)\mathcal{O}_{p'}(0)\rangle =\delta_{p,p'}$, and the OPE coefficients are real in this basis. $\mathcal{O}_p$ obeys the unitarity bound, which is that $\Delta_p \geq l_p+d-2$ for $l_p>0$, and $\Delta_p \geq (d-2)/2$ for $l_p=0$. For identical external operators, conformal blocks are independent of the external operator dimension.  For $z=\bar{z}$, the blocks behave as
\begin{equation}
g_{\Delta_p,l_p}(z\rightarrow 0) \sim z^{\Delta_p} ~~~~~~g_{\Delta_p,l_p}(z\rightarrow 1 ) \sim \log^2(1-z).
\end{equation}
A finite number of t-channel blocks cannot reproduce the $u^{-\Delta}$ s-channel OPE singularity, and so the s-channel singularity must be reproduced by an infinite sum in the t-channel. In even dimensions, the blocks are known in closed form. We will often study two dimensions as an example, for which the blocks are
\begin{equation}
g_{\Delta,l}(z, \bar{z}) =
\frac{k_{\Delta+l}(z) k_{\Delta-l}(\bar{z}) + k_{\Delta-l}(z) k_{\Delta+l}(\bar{z})}{1+\delta_{0,l}},
\label{2dBlock}
\end{equation}
where $k_\beta(z) = z^{\beta/2} \F(\beta/2, \beta/2, \beta;z)$.

The correlator can also be written as a sum over states in terms of scaling blocks and coefficients $a_{h',\bar{h}'} \geq 0$ as
\begin{equation}
G(z,\bar{z}) = \sum_{h', \bar{h}'} a_{h',\bar{h}'} z^{h'} \bar{z}^{\bar{h}'},
\end{equation}
where $h' = \frac{1}{2}(\Delta' \pm l')$, $\bar{h}' = \frac{1}{2}(\Delta' \mp l')$ \cite{HartmanJK15}. The sum runs over primaries and descendants. For simplicity, we will mainly work with the scaling-block decomposition.

\section{Light-state dominance for $z = \bar{z}$}
\label{sec:main_result}

In this section we will prove the essential result of this work, light-state dominance in the kinematic regime $z=\bar{z}$ real with $z \in [0,1]$. We will show that there exists a channel in which the total contribution of all exchanged operators $\mathcal{O}_p$ with dimension $\Delta_p > \Delta_c > 2\Delta$ is exponentially suppressed in the cutoff $\Delta_c$ relative to the contribution from operators with $\Delta_p < \Delta_c$. Subsequent sections are extensions and applications of this result, so while the proof is elementary, we will go through the steps in detail.

Our proof is an application of the HKS modular invariance argument for the partition function \cite{HartmanKS14}. Our presentation most closely follows the review in \cite{KrausSS17}. We will divide the spectrum into heavy and light according to cutoffs $h_c + \bar{h}_c \equiv \Delta_c$. We will often use simplified notation $ G^s \equiv G(z,\bar{z})$ and $G^t \equiv G(1-z,1-\bar{z})$. The correlator is divided into heavy and light exchanged operator contributions in each channel as
\begin{equation}
G^s = G^s_L + G^s_H,
~~~
G^t = G^t_L + G^t_H.
\end{equation}
The crossing equation can be written as
\begin{equation}
v^\Delta G^s_L - u^\Delta  G^t_L = u^\Delta  G^t_H - v^\Delta G^s_H.
\label{KSS2.5}
\end{equation}
One way to understand this statement of crossing is that under the crossing transformation the light contribution gains what the heavy contribution loses.

\subsection{Light-state dominance with scaling blocks}

We will first prove light-state dominance  using scaling blocks. The correlator is
\begin{equation}
G(z) = \sum_{h', \bar{h}'} a_{h', \bar{h}'} z^{\Delta'}.
\end{equation}
We first consider $z<1/2$. We begin with the bound
\begin{equation}
G^s_H
=
\sum_{h'>h_c, \bar{h}'>\bar{h}_c} a_{h', \bar{h}'} z^{\Delta'}
=
\sum_{h'>h_c, \bar{h}'>\bar{h}_c} a_{h', \bar{h}'} \left( \frac{z}{1-z} \right)^{\Delta'} (1-z)^{\Delta'}
\leq
\left( \frac{z}{1-z} \right)^{\Delta_c} G^t_H.
\end{equation}
Using $v^\Delta = (v/u)^\Delta u^\Delta,$ we have
\begin{equation}
v^\Delta G^s_H
\leq
R u^\Delta G^t_H,
\label{KSS2.6}
\end{equation}
with
\begin{equation}
R = \left( \frac{z}{1-z} \right)^{\Delta_c-2\Delta}~.
\end{equation}
Now that an upper bound \eqref{KSS2.6} has been established, the rest of the proof   proceeds as in the partition function case \cite{HartmanKS14}. Subtracting $u^\Delta G^t_H$ from both sides,
\begin{equation}
v^\Delta G^s_H-u^\Delta  G^t_H
\leq
(R-1)
u^\Delta G^t_H.
\label{KSS2.7}
\end{equation}
 We will take $\Delta_c > 2\Delta$ so that $R < 1$, which in turn implies that
\begin{equation}
u^\Delta G^t_H
\leq
\frac{1}{1-R}
\left(
u^\Delta  G^t_H-v^\Delta G^s_H
\right).
\end{equation}
Using \eqref{KSS2.6}, we obtain a bound on the heavy-state contribution in terms of heavy data,
\begin{equation}
v^\Delta G^s_H
\leq
\frac{R}{1-R}
\left(
u^\Delta  G^t_H-v^\Delta G^s_H
\right).
\end{equation}
Using crossing symmetry \eqref{KSS2.5}, the bound on the heavy state contribution is now in terms of light data.
\begin{equation}
v^\Delta G^s_H
\leq
\frac{R}{1-R}
\left(
v^\Delta  G^s_L -u^\Delta G^t_L\right).
\label{HeavySChannelStrongBound}
\end{equation}
This is the strongest bound on the heavy-state contribution we will derive. One way to understand \eqref{HeavySChannelStrongBound} is that crossing mandates that the density and OPE coefficients of heavy states cannot be too large. A remarkable consequence is that the contribution from all heavy operators is bounded by the light data, even though in a general strongly-coupled CFT, analytic bootstrap techniques access only certain families of heavy operators (for example \cite{FitzpatrickKPS12,KomargodskiZ12,OPEInversion,LargeSpinPerturbationTheory,Simmons-Duffin16}), or contributions with asymptotically high energies \cite{PappadopuloRER12,RychkovY15,MukhametzhanovZ18}. One way to understand this difference is that unlike previous bootstrap approaches, the constraints we derive are not explicitly based on the singularity structure of the crossing equation.

It will be convenient to bound the heavy s-channel contribution in terms of the light s-channel contribution, which we achieve by using positivity to drop the second term in \eqref{HeavySChannelStrongBound}.
\begin{equation}
G^s_H
\leq
\frac{R}{1-R}
G^s_L .
\label{HeavySChannelBound}
\end{equation}
This translates into a bound on the correlator,
\begin{equation}
G^s_L \leq G^s = G^s_L+ G^s_H \leq \frac{1}{1-R} G^s_L.
\end{equation}
The analogous bound for $z>1/2$ is derived similarly. For $z>1/2$,
\begin{equation}
u^\Delta G^t_H \leq  R^{-1} v^{\Delta} G^s_H,
\end{equation}
where $R^{-1} < 1$. From here the steps proceed as before. The bound on the heavy-state contribution in the s-channel is
\begin{equation}
\frac{1}{1-R^{-1}} \left( u^\Delta G_L^t - v^\Delta G_L^s \right) \geq v^\Delta G_H^s.
\label{HeavySChannelBoundLargeZ}
\end{equation}
Carrying through the remainder of the argument, we arrive at
\begin{equation}
G^t_L \leq G^t \leq \frac{1}{1-R^{-1}} G^t_L.
\end{equation}
To summarize, we have shown that
\begin{equation}
G(z) =
\left\{
\begin{array}{ll}
 ~~~~~~~~~~~G_L(z) \left(1+\mathcal{E}(z) \right) ,& z < 1/2, \\
\left(\frac{z}{1-z} \right)^{2\Delta}  G_L(1-z) \left(1+\mathcal{E}(1-z)\right) , & z > 1/2,
\end{array}  \right.
\label{LightStateBoundZZbarEqual}
\end{equation}
where the contribution from heavy states is bounded according to $ \mathcal{E}(z)  \leq \frac{R(z)}{1-R(z)}$ and $R(z) = \left(\frac{z}{1-z} \right)^{\Delta_c-2\Delta}$.  The validity of this bound requires $\Delta_c > 2\Delta$. As long as $z, \bar{z}< 1/2$, or $z,\bar{z}>1/2$, we may take $z \neq \bar{z}$ and the derivation proceeds in the same way. In this work, weaker upper bounds of the form \eqref{LightStateBoundZZbarEqual} will often be sufficient for our purposes, but it should always be understood that the strongest bounds arise from \eqref{HeavySChannelStrongBound}, \eqref{HeavySChannelBoundLargeZ}. The difference is most apparent near $z=1/2$, as we will examine in detail later.

It is straightforward to verify these bounds using explicit correlators. For example, expanding the mean field theory correlator
\begin{equation}
G_{MFT}(z) = 1+z^{2\Delta}+\left(\frac{z}{1-z}\right)^{2\Delta}
\end{equation}
about $z = 0, 1$ gives the scaling block decomposition used in this section. One can check that this correlator does not saturate either the weak or strong version of the  bounds for all values of $z$, and we have no reason to expect that the strong bound can be saturated by any correlator for all $z$.

We have bounded $G_H$ relative to $G_L$, and so refining our knowledge of $G_L$ is important for placing stronger bounds on the correlator. $G_L(z)$ for $z<1/2$  can be bounded by a function that is independent of $\Delta_c$,
\begin{equation}\label{sparse}
G_L(z) \leq f_s (z), ~~~~~~~~~~~~~~~~~~~~~~~~~\frac{d}{d \Delta_c} f_s(z) = 0,
\end{equation}
which implies the analogous condition $G_L(1-z) \leq f_s(1-z)$ for $z > 1/2$.  For example, this is satisfied by taking $f_s(z)$ to be the full correlator $G(z)$.    We refer to (\ref{sparse}) as a sparseness condition, since if the light spectrum is sufficiently sparse we can choose an $f_s(z)$ that reflects this. The optimal choice of $f_s$ is theory-dependent.  It is however true that because the correlator is bounded away from its OPE singularities, there exists some constant $A_s>0$ such that
\begin{equation}
f_s(z) = A_s G_{MFT}(z),
\label{Sparseness}
\end{equation}
is a valid bound for any correlator \footnote{As $G(z)$ is finite away from its OPE singularities, there exists a $B(\epsilon)$ such that $G(z) < B(\epsilon)$ for $z \in [\epsilon,1-\epsilon]$. Here $\epsilon \ll 1$. There also exists an $\epsilon_0$ such that for $z < \epsilon_0$, $G(z)$ is dominated by leading $z=0$ term (which is 1, as the factor of $u^{-\Delta}$ has been stripped off). Therefore, choose $A_s = B+\delta$ for some $0< \delta<1$.} The mean field theory correlator admits $A_s=1$ \footnote{An even more stringent upper bound comes from using only the light contribution to $G_{MFT}$, but this would explicitly depend on $\Delta_c$, albeit in a known way.}. Using the sparseness condition \eqref{Sparseness} or any other, we have
\begin{equation}
G(z) =
\left\{
\begin{array}{ll}
 ~~~~~~~~~~~~~G_L(z) +\mathcal{O}\left(  \left(\frac{z}{1-z} \right)^{\Delta_c-2\Delta}  \right) ,& z < 1/2, \\
\left(\frac{z}{1-z} \right)^{2\Delta} \left( G_L(1-z) +\mathcal{O}\left(  \left(\frac{1-z}{z} \right)^{\Delta_c-2\Delta}  \right) \right),& z > 1/2,
\end{array}  \right.
\label{SparseLightStateBoundZZbarEqual}
\end{equation}
where the notation $\mathcal{O}\left( \left(\frac{z}{1-z} \right)^{\Delta_c-2\Delta}  \right)$ refers to the large-$\Delta_c$ decay rate at fixed $z$. The correlator is therefore well-approximated by the contribution from light states with a well-controlled error.

When there are two energy scales in the theory, light-state dominance provides additional information about the relative suppression of high-energy effects.
For example, consider correlators of single-trace operators with $\Delta \sim \mathcal{O}(N^0)$, which by definition have an asymptotic expansion in $1/N$. Choosing $\Delta_c \sim N$ shows contributions to generalized free correlators that are non-perturbative in $N$ are suppressed as $e^{-N}$, while the contributions of massive string states can be as large as $e^{-\Delta_{gap}} \gg e^{-N}$ in theories with $\Delta_{gap} \ll N$. For these correlators, such a hierarchy of the $\Delta_{gap}$ and $N$ in the CFT is dual to a bulk theory in which stringy effects dominate loop effects.

\subsection{Light-state dominance with conformal blocks}

In this section, we will use the conformal block expansion of the correlator and divide heavy and light according to the weights of primaries. We will derive a bound analogous to \eqref{LightStateBoundZZbarEqual} that will end up being stronger. We will focus on two dimensions as a representative example and use the explicit form of the conformal blocks, and we will also comment on the proof in arbitrary dimensions along the way. We begin with the case $z<1/2$.
\begin{equation}
G^s_H = \sum_{h_p > h_c, \bar{h}_p> \bar{h}_c} C_{h_p, \bar{h}_p}^2 z^{\Delta_p}\F(h_p,h_p,2h_p,z)\F(\bar{h}_p,\bar{h}_p,2\bar{h}_p,z),
\end{equation}
where $C_{h_p, \bar{h}_p}^2$ are the squares of OPE coefficients of primaries.  Consider the ratio
\begin{equation}
r=\frac{\F(h,h,2h,z)}{\F(h,h,2h,1-z)}~,\quad z<{1\over 2}
\label{RatioOf2F1s}
\end{equation}
It is a fact that $r \leq 1$.   This follows by noting that the hypergeometric functions have a convergent expansion with positive coefficients, so each term in the numerator is smaller than the corresponding term in the denominator. The same argument shows that $g_{\Delta_p,l_p}(z) \leq g_{\Delta_p,l_p}(1-z)$ in arbitrary dimensions as well.

One can check numerically that $r$ decays monotonically in $h$. We have checked the monotonic decay of the ratio of blocks in four dimensions as well.  We expect this property to hold for blocks in all dimensions, and it would be interesting to obtain an analytic proof. We now have a bound that is analogous to \eqref{KSS2.6},
\begin{equation}
v^\Delta G^s_H
\leq
R u^\Delta G^t_H,
\end{equation}
where
\begin{equation}
R = \frac{v^\Delta g_{h_c,\bar{h}_c}(u,v)}{u^\Delta g_{h_c,\bar{h}_c}(v,u)} = \left(\frac{z}{1-z}\right)^{\Delta_c-2\Delta} \frac{\F(h_c,h_c,2h_c,z)\F(\bar{h}_c,\bar{h}_c,2\bar{h}_c,z)}{\F(h_c,h_c,2h_c,1-z)\F(\bar{h}_c,\bar{h}_c,2\bar{h}_c,1-z)} \leq 1,
\label{RforBlocks}
\end{equation}
as long as $\Delta_c > 2\Delta $. To extend the bound to arbitrary dimensions, we can instead use $g_{\Delta_p,l_p}(z) \leq g_{\Delta_p,l_p}(1-z)$, in which case $R = (u/v)^{\Delta_c-2\Delta}$. The remaining steps proceed as in the scaling block case. Light-state dominance therefore applies to the contributions of light and heavy primaries.

The ratio of hypergeometric functions provides an additional suppression over the scaling block result. In fact, one can show that the additional suppression provided by the conformal block leads to $R<1$ for $\Delta_c$ in a certain window below $2\Delta$, in contrast to the scaling block case, but only for values in a restricted region $[\epsilon,1/2]$. As $\Delta_c \rightarrow 2\Delta$, $\epsilon \rightarrow 0$.

\subsection{Relation to estimates from OPE convergence}

We have derived bounds on the heavy-state contribution for any $\Delta_c >2\Delta$. We will compare this bound to the results in \cite{PappadopuloRER12,RychkovY15, MukhametzhanovZ18}, which use OPE behavior to estimate the contribution of states with $\Delta_c \gg \Delta $.

First, we briefly review the bounds derived in \cite{PappadopuloRER12}. The analysis considered $1-z < z_0 \ll 1$ with $z_0$ small enough so that the t-channel OPE singularity is a good approximation to the correlator. $z_0$ is theory dependent, as for example large light OPE coefficients imply small $z_0$. On the cylinder, $G(z)$ takes the form of a Laplace transform of s-channel OPE data, which must reproduce the t-channel OPE singularity when $z_0 \ll 1$. The inverse Laplace transform of the t-channel OPE singularity therefore gives the $\Delta' \rightarrow \infty$ asymptotics of the s-channel OPE spectral density $f(\Delta')$. The Hardy-Littlewood tauberian theorem controls the error in $f(\Delta')$ for $\Delta' >\Delta_{HL}$, the so-called Hardy-Littlewood threshold, that is incurred by dropping subleading t-channel singularities. The quantities $z_0 $ and $\Delta_{HL}$ encode similar physics, as the smaller the window in which the leading OPE singularity dominates the correlator, the less the leading singularity constrains the contribution of the high-energy states. In particular, $\Delta_{HL} \sim \Delta/z_0$. The asymptotic behavior of $f(\Delta')$ translates into a bound on the contribution from operators above dimension $\Delta_c \gg  \Delta_{HL}$ to the correlator. This heavy-state contribution decays exponentially as $\Delta_c \rightarrow \infty$, establishing exponentially-fast convergence of the s-channel OPE. The decay estimate was improved in \cite{RychkovY15} and subleading corrections were calculated in \cite{MukhametzhanovZ18}, in which this approach was extended.

Our heavy-state bound shares some similarities with the estimates in \cite{PappadopuloRER12}. Like the OPE convergence estimate in which $\Delta_{HL}$ is theory-dependent, the fact that $A_s$ is unknown means that there exists some $\Delta_c$ such that the heavy-state contribution is exponentially small, but this value is theory-dependent. As such, our heavy state bound may provide an alternative understanding for the existence of a theory-dependent threshold $\Delta_{HL}$. The value of $\Delta_{HL}$ is $z$-dependent, as is the optimal choice of $f_s(z)$ and therefore the smallest possible value for $A_s$.

Unlike the OPE convergence estimate however, our heavy-state upper bound decays exponentially in $\Delta_c$ for all $\Delta_c > 2\Delta$ even though the true heavy-state contribution may not decay exponentially until $\Delta_c \sim \Delta_{HL}$. When a sparseness condition is imposed that determines how $G_L$ depends on $\Delta_c$, our heavy-state bounds are exact bounds on the heavy state contribution above $2\Delta$, while the estimates derived from OPE convergence and corrected by including subleading singularities remain asymptotic, and computing $\Delta_{HL}$ may be less straightforward \cite{MukhametzhanovZ18}.

Having understood the ranges of validity, we will compare the OPE convergence estimates and heavy-state bounds quantitatively. The strongest OPE convergence bound derived in \cite{HogervorstR13,RychkovY15} for the $\rho$ coordinate \eqref{RhoCoordinates} is
\begin{equation}
\sum_{\Delta_p>\Delta_c} a_{\Delta_p} g_{h_p,\bar{h}_p}(\rho)
\leq
\frac{2^{4\Delta+1} \Delta_c^{4\Delta-1}}{\Gamma(4\Delta)}
\frac{\rho^{\Delta_c}}{1-\rho^2}
\label{RychkovBound}
\end{equation}
for $\Delta_c \gg \Delta_{HL}, \Delta_c \gg 2\Delta/t $ where $\rho= e^{-t}$. We have restricted $\rho = \bar{\rho}$ real.

First we consider $z<1/2$. Changing coordinates, $\frac{z}{1-z}= \frac{4\rho}{(1-\rho)^2} $, our  scaling block bound is
\begin{equation}
\sum_{\Delta_p>\Delta_c} a_{\Delta_p} g_{h_p,\bar{h}_p}(z) \leq
\sum_{\Delta'>\Delta_c} a_{\Delta'} z^{\Delta'}
\leq
G_L(z) \left( \frac{4\rho}{(1-\rho)^2} \right)^{\Delta_c-2\Delta}
\left(
1-\left( \frac{4\rho}{(1-\rho)^2} \right)^{\Delta_c-2\Delta}
\right)^{-1}
,
\label{LSUComparison}
\end{equation}
for $\Delta_c > 2\Delta$. The upper bound \eqref{Sparseness} on $G_L(z)$ is independent of $\Delta_c$. The heavy-state bound has the same leading exponential dependence $\rho^{\Delta_c}$ as \eqref{RychkovBound} but including the subleading terms we see that the heavy-state bound is slightly weaker. The bound \eqref{LSUComparison} becomes dramatically weaker as $z \rightarrow 1/2$. The behavior near $z=1/2$ is more effectively bounded by the stronger bound \eqref{HeavySChannelStrongBound} . The strong upper bound's dependence on $\Delta_c$ at $z=1/2$ is $1/(\Delta_c-2\Delta)$ from identity exchange, and by including additional states the decay in $\Delta_c$ can only become weaker. This is weaker than the suppression $\Delta_c^{4\Delta-1} (3-2\sqrt{2})^{\Delta_c}$ in \eqref{RychkovBound}. For $z > 1/2$, the strong heavy state bound is \eqref{HeavySChannelBoundLargeZ}
\begin{equation}
v^\Delta G_H(z) \leq \frac{1}{1-((1-z)/z)^{\Delta_c-2\Delta}} (u^\Delta G_L(1-z)-v^\Delta G_L(z)),
\label{LSUComparisonNear1}
\end{equation}
which again decays at best as $1/(\Delta_c-2\Delta)$, which is due to identity exchange.

The heavy state bound that uses conformal blocks  is stronger than the scaling block bound due to the additional suppression provided by the ratio of conformal blocks \eqref{RforBlocks}. This should be compared to the analogous OPE convergence estimate derived from the conformal block expansion.

Notice that the OPE convergence estimate \eqref{RychkovBound} and heavy-state bound \eqref{LSUComparisonNear1} on $G_H(\rho)$ are singular as $\rho \rightarrow 1$ for fixed $\Delta_c$. This is necessary, because the heavy states exchanged in the s-channel must reproduce the t-channel singularity, and so any upper bound on these heavy states must become infinite in the t-channel OPE limit.

We have shown that the strongest heavy-state bound is weaker than the OPE convergence estimate \eqref{RychkovBound}. As discussed, depending how much is known about the light data, the heavy-state bound may be valid in a wider range of $\Delta_c$. These conclusions were inevitable outcomes. The estimate \eqref{RychkovBound} uses the OPE singularity structure, which places a stronger constraint on the high-energy behavior than our application of crossing symmetry, but unlike our method, using the OPE singularity constrains only asymptotically high energies. The difference between the OPE convergence estimate \eqref{RychkovBound} and light-state dominance is analogous to the difference between the Cardy formula and the HKS result respectively. The analogy is not superficial, but occurs at the level of technical similarities between the derivations of the partition-function and correlator results.

\section{Light-state dominance on the first sheet}
\label{sec:first_sheet}

In this section, we extend light-state dominance to all $z, \bar{z}$ on the first sheet. Light state contributions will determine the correlator up to bounded corrections. The heavy-state bounds take essentially the same form as in section \ref{sec:main_result}, relative exponential suppression in $h_c, \bar{h}_c$, but the exact bounds will depend on the kinematics in a variety of ways. We will sometimes prove results for the partition function as a warmup, as the procedure is much the same for the correlator. For simplicity, we will derive bounds using scaling blocks rather than conformal blocks, but as previously illustrated, the conformal block bounds follow without any fundamentally new ingredients. Although we will not do so here, it would be interesting to study light-state dominance in Lorentzian regimes off the first sheet and determine at what times light-state dominance breaks down.

\subsection{Revisiting Z: $\beta_L > 2\pi, \beta_R< 2\pi$}

Before we derive light-state dominance for $z<1/2, \bar{z} > 1/2,$ we will prove the analogous statement for the partition function with $\beta_L > 2\pi , \beta_R < 2\pi$. We follow \cite{HartmanKS14}, in which $\beta' = 4 \pi^2/\beta$ under modular transformation and the cutoff for the heavy states is placed at energy $\epsilon> 0$. The heavy contribution $Z_H$ to the partition function $Z$ is
\begin{equation}
Z_H(\beta_L, \beta_R) = \sum_{E_L > \epsilon_L, E_R > \epsilon_R} \rho(E_L,E_R) e^{-\beta_L E_L - \beta_R E_R}
\leq
e^{-\epsilon_L(\beta_L-\beta_L')} Z_H(\beta_L', \beta_R).
\end{equation}
We will use $R \equiv e^{-\epsilon_L(\beta_L-\beta_L')} < 1$. Applying the familiar HKS argument, we arrive at
\begin{equation}
Z_H(\beta_L, \beta_R) \leq \frac{R}{1-R}(Z_H(\beta_L', \beta_R)-Z_H(\beta_L, \beta_R)) .
\end{equation}
The relevant modular invariance condition is
\begin{equation}
Z_L(\beta_L', \beta_R) + Z_H(\beta_L', \beta_R)
=
Z_L(\beta_L, \beta_R') + Z_H(\beta_L, \beta_R').
\end{equation}
Using this condition, we have
\begin{equation}
Z_H(\beta_L, \beta_R) \leq \frac{R}{1-R}(Z_H(\beta_L, \beta_R')-Z_H(\beta_L, \beta_R)
+Z_L(\beta_L, \beta_R')-Z_L(\beta_L', \beta_R)).
\end{equation}
Unlike the $\beta_L = \beta_R$ case, the first two terms do not cancel, which will make our bound in some sense weaker. Due to positivity,
\begin{equation}
Z_H(\beta_L, \beta_R) \leq \frac{R}{1-R}(Z_H(\beta_L, \beta_R')
+Z_L(\beta_L, \beta_R')).
\end{equation}
We can bound the term $Z_H(\beta_L, \beta_R')$ that was not present in the $\beta_L = \beta_R$ argument. The original partition function bound in \cite{HartmanKS14} applies to $\beta_L, \beta_R' > 2\pi$ without any real modification due to $\beta_L \neq \beta_R'$. The bound in \cite{HartmanKS14} is
\begin{equation}
Z_H(\beta) \leq \frac{e^{\epsilon(\beta'-\beta)}}{1-e^{\epsilon(\beta'-\beta)}} Z_L(\beta).
\end{equation}
Using this bound,
\begin{equation}
Z_H(\beta_L, \beta_R') \leq \frac{e^{\epsilon_L(\beta_L'-\beta_L)+\epsilon_R(\beta_R-\beta_R')}}{1-e^{\epsilon_L(\beta_L'-\beta_L)+\epsilon_R(\beta_R-\beta_R')}} Z_L(\beta_L, \beta_R')
\equiv \frac{R'}{1-R'} Z_L(\beta_L, \beta_R').
\end{equation}
We have now bounded the heavy contribution in terms of light data, our original goal,
\begin{equation}
Z_H(\beta_L, \beta_R) \leq \frac{R}{(1-R)(1-R')}Z_L(\beta_L, \beta_R').
\end{equation}
We then have
\begin{equation}
Z_L(\beta_L, \beta_R) \leq Z(\beta_L, \beta_R) \leq Z_L(\beta_L, \beta_R) + \frac{R}{(1-R)(1-R')}Z_L(\beta_L, \beta_R').
\end{equation}
In summary, the free energy is
\begin{equation}
\ln Z(\beta_L, \beta_R) \leq \ln Z_L(\beta_L,\beta_R)+
\ln \left(
1+ \frac{R }{(1-R)(1-R')}\frac{Z_L(\beta_L, \beta_R')}{Z_L(\beta_L, \beta_R)}
\right).
\end{equation}
The bound is effective only if $R'$ is small enough compared to $R$, but for large enough $\epsilon$, $\frac{R }{(1-R)(1-R')} \ll 1$. We will defer detailed analysis of the finite-$\epsilon$ behavior until next section, in which the analogous criterion will arise for the correlator.

Light-state dominance of the free energy is of interest for large-$c$ theories. Unlike in the $\beta_L = \beta_R$ case, the error term here is not automatically $\mathcal{O}(c^0)$ because  the ratio of partition functions can grow with $c$. Further assuming a sub-Hagedorn density of left and right-moving light states, the vacuum dominates, and the error term grows with $c$ approximately as $\ln(1 + e^{(\beta_R'-\beta_R) c/24}) \approx (\beta_R'-\beta_R)c/24$. As expected, the regime $\beta_L> 2\pi, \beta_R < 2\pi$ is probing the light sector of two different modular frames, and so the free energy is not always well-approximated by the free energy in the original modular frame. Physically, this is expected, as it corresponds to competition between thermal AdS and BTZ free energies at large $c$. We will not further explore this or related results for $\beta_L \neq \beta_R$, as these are only warm-ups for our main focus, the correlator. For a much more complete treatment of $\beta_L \neq \beta_R$, see \cite{HartmanKS14,AnousMS18}.

\subsection{Light-state dominance for $z<1/2, \bar{z} > 1/2$}

We now show that for $z<1/2, \bar{z} > 1/2$ real, the correlator is determined by light data up to bounded corrections. We begin with the bound
\begin{equation}
G_H(z, \bar{z})  \leq \left(\frac{z}{1-z}\right)^{h_c} \sum_{h' \geq h_c,\bar{h}' \geq \bar{h}_c} a_{h',\bar{h}'} (1-z)^{h'} \bar{z}^{\bar{h}'}
=\left(\frac{z}{1-z}\right)^{h_c} G_H(1-z,\bar{z}),
\end{equation}
where we have dropped the superscripts $s,t$ as they are no longer meaningful. It follows that
\begin{equation}
G_H(z, \bar{z})  \leq \left(\frac{z}{1-z}\right)^{h_c} G_H(1-z,\bar{z}),
\end{equation}
Using $R \equiv \left(\frac{z}{1-z}\right)^{h_c}$, the familiar steps lead us to
\begin{equation}
G_H(z,\bar{z})   \leq \frac{R}{1-R} ( G_H(1-z,\bar{z})-G_H(z, \bar{z})   ).
\end{equation}
Using the crossing relation
\begin{equation}
(z (1-\bar{z}))^\Delta(G_L(1-z, \bar{z}) + G_H(1-z, \bar{z}))
=
((1-z)\bar{z})^\Delta (G_L(z, 1-\bar{z}) + G_H(z, 1-\bar{z})),
\end{equation}
we have
\begin{align*}
G_H(z,\bar{z})   &\leq \frac{R}{1-R} \bigg(
\left(\frac{(1-z)\bar{z}}{z (1-\bar{z})}\right)^\Delta (G_L(z, 1-\bar{z}) + G_H(z, 1-\bar{z}))
\\
&~~~~~~~~
-G_L(1-z, \bar{z})
-G_H(z, \bar{z})\bigg).
\end{align*}
Dropping terms due to positivity,
\begin{equation}
G_H(z,\bar{z})   \leq \frac{R}{1-R} \left(
\left(\frac{(1-z)\bar{z}}{z (1-\bar{z})}\right)^\Delta (G_L(z, 1-\bar{z}) + G_H(z, 1-\bar{z}))
\right).
\end{equation}
The second term we have already bounded in \eqref{HeavySChannelBound},
\begin{equation}
G_H(z, 1-\bar{z}) \leq \frac{R'}{1-R'}G_L(z,1-\bar{z}),
\end{equation}
where $R' = \left(\frac{z(1-\bar{z})}{(1-z)\bar{z}}\right)^{\Delta_c-2\Delta}$. We therefore obtain the bound
\begin{equation}
G_H(z,\bar{z})   \leq \frac{R}{(1-R)(1-R')} \left(
\left(\frac{(1-z)\bar{z}}{z (1-\bar{z})}\right)^\Delta G_L(z, 1-\bar{z})
\right).
\end{equation}
As we are using scaling blocks, it is clear that $G_L(z, 1-\bar{z}) \leq G_L(z, \bar{z})$. One can check that this is true using the conformal block expansion as well. For example, in 2d the blocks are proportional to $\F(k,k,2k,z)$, which increases monotonically in $z$. The bound becomes
\begin{equation}
G_H(z,\bar{z})   \leq \frac{R}{(1-R)(1-R')} \left(
\left(\frac{(1-z)\bar{z}}{z (1-\bar{z})}\right)^\Delta G_L(z, \bar{z})
\right).
\end{equation}
The correlator is well-approximated by the s-channel light-state contribution. The requirement $R' < 1$ is satisfied for $z, 1-\bar{z} < 1/2$. The bound on the correlator is
\begin{equation}
G_L(z, \bar{z}) \leq G(z, \bar{z}) \leq G_L(z, \bar{z})
\left( 1 +  \frac{R}{R''(1-R)(1-R')} \right) ,
\label{LowTempZZbarDifferent}
\end{equation}
where we have defined $R'' = \left(\frac{z (1-\bar{z})}{(1-z)\bar{z}}\right)^\Delta \leq 1$. The error term $\frac{R}{R''(1-R)(1-R')} $ decays exponentially in $h_c$ for sufficiently large $h_c$.

The error term is controlled by the ratio $R/R''$, which can be greater than 1 for some values of $z,\bar{z}$. The condition $R/R'' < 1$ is
\begin{equation}
\left( \frac{z}{1-z} \right)^{h_c-\Delta} < \left( \frac{1-\bar{z}}{\bar{z}}\right)^{\Delta}.
\end{equation}
For $h_c > 2\Delta$, the inequality is satisfied for all $z<1-\bar{z}<1/2$ but not in the entire complementary region $1-\bar{z} < z< 1/2$.

We now prove a bound for the region $1-\bar{z} < z < 1/2$. The bound \eqref{LowTempZZbarDifferent} was derived using a modification of the $z, \bar{z} < 1/2$ proof. We can modify the procedure for the $z,\bar{z} > 1/2$ case in order to bound the correlator according to its $\bar{z}>1/2$ value rather than $z<1/2$ value, and it is obvious that this will bound the correlator in the remaining region. The relevant bound is
\begin{equation}
G_H(1-z,1-\bar{z}) \leq \left(\frac{1-\bar{z}}{\bar{z}}\right)^{\bar{h}_c} \sum_{h' \geq h_c,\bar{h}' \geq \bar{h}_c} a_{h',\bar{h}'} (1-z)^{h'} \bar{z}^{\bar{h}'}
=\left(\frac{1-\bar{z}}{\bar{z}}\right)^{\bar{h}_c} G_H(1-z,\bar{z}).
\end{equation}
The relevant small quantity is $R \equiv \left(\frac{1-\bar{z}}{\bar{z}}\right)^{\bar{h}_c}$. Following familiar steps we arrive at
\begin{equation}
G_H(1-z,1-\bar{z})   \leq \frac{R}{1-R} ( G_H(1-z,\bar{z})-G_H(1-z,1-\bar{z})  ).
\end{equation}
Using crossing and dropping negative terms, we have
\begin{equation}
G_H(1-z,1-\bar{z})   \leq \frac{R}{1-R}
\left( \frac{(1-z)\bar{z}}{z (1-\bar{z})}\right)^\Delta (G_L(z, 1-\bar{z}) + G_H(z, 1-\bar{z})).
\end{equation}
As before, we have already bounded $G_H(z, 1-\bar{z})$ and so
\begin{equation}
G_H(1-z,1-\bar{z})  \leq \frac{R}{R''(1-R)(1-R')}  G_L(z, 1-\bar{z}).
\end{equation}
At this point, we may use either $G_L(z, 1-\bar{z}) \leq G_L(1-z, 1-\bar{z})$ or $G_L(z, 1-\bar{z}) \leq G_L(z, \bar{z})$ to reformulate the bound in terms of s or t-channel data. Using crossing symmetry and $G_L(z,1-\bar{z}) < G_L(1-z,1-\bar{z})$, we have
\begin{equation}
\left( \frac{u}{v} \right)^{\Delta }G_L(1-z,1-\bar{z}) \leq G(z,\bar{z})  \leq G_L(1-z, 1-\bar{z}) \left( \frac{u}{v} \right)^{\Delta }
\left(1+
\frac{R}{R''(1-R)(1-R')}
\right)
.
\label{HighTempZZbarDifferent}
\end{equation}
In this bound, $R/R''< 1$ if
\begin{equation}
\left( \frac{1-\bar{z}}{\bar{z}}\right)^{\bar{h}_c} \leq \left( \frac{z}{1-z}\right)^\Delta.
\end{equation}
For $\bar{h}_c>2\Delta$, the inequality is satisfied for all $1-\bar{z} < z < 1/2$, giving us bounds that cover the full range of $z, 1-\bar{z} < 1/2$ as desired. Bounds for the case of $z>1/2, \bar{z}<1/2$ follow similarly.

\subsection{Light-state dominance for complex $z, \bar{z}$}

The bounds we have derived for real $z, \bar{z}$ can be extended to complex values everywhere the s, t-channel OPEs both converge.  By expanding in $(\rho,\overline{\rho})$ rather than $(z,\overline{z})$, where
\begin{equation}
z = \frac{4 \rho}{(1+\rho)^2}~~~~~~~~~ \rho = \frac{1-\sqrt{1-z}}{1+\sqrt{1-z}},
\label{RhoCoordinates}
\end{equation}
one achieves convergence everywhere on the $z$-plane
except for a branch cut between $1$ and $\infty$ \cite{PappadopuloRER12,HogervorstR13}.
The complex $z$ plane is mapped to the interior of the $|\rho| < 1$ disk and the branch cut is mapped to the boundary of the disk. The crossing-symmetric point $z = 1/2$ maps to $\rho = 3 - 2 \sqrt{2} \approx 0.17$. The crossing transformation is $\rho \rightarrow \rho' \equiv \left( \frac{1-\sqrt{\rho}}{1+\sqrt{\rho}} \right)^2$. The ratio $|\rho/\rho'| < 1$ for $|\rho| < 3 - 2 \sqrt{2}$. The correlator can be decomposed into $\rho, \bar{\rho}$ scaling blocks with positive coefficients \cite{HartmanJK15} as 
\begin{equation}
G(\rho,\bar{\rho}) =
\sum b_{h_p,\bar{h}_p} \rho^{h_p} \bar{\rho}^{\bar{h}_p},
\end{equation}
in which crossing is $(16 \rho \bar{\rho})^{-\Delta} G(\rho, \bar{\rho}) = (16 \rho' \bar{\rho}')^{-\Delta} G(\rho', \bar{\rho}')$. It now follows that our earlier derivations of light-state dominance go through in $(\rho, \bar{\rho})$ coordinates as well.

The heavy-state contribution for complex $(\rho, \bar{\rho})$ can be obtained by analytic continuation. The continuation simply adds phases to each term, and so
\begin{equation}
|G_H(\rho, \bar{\rho})| \leq
\sum b_{h_p,\bar{h}_p} |\rho|^{h_p} |\bar{\rho}|^{\bar{h}_p} \leq G_H(|\rho|, |\bar{\rho}|),
\end{equation}
which we have already bounded. The phases can only decrease the magnitude of the heavy-state contribution, proving a bound on the heavy contribution everywhere on the first sheet.

The statement here is that the heavy-state contribution with complex $(\rho, \bar{\rho})$ is bounded by the light-state contribution with real $(\rho, \bar{\rho})$. For complex $(\rho, \bar{\rho})$, our bounds no longer tie the size of the heavy-state contribution to the light-state contribution at the same $(\rho, \bar{\rho})$. Indeed, for finely-tuned kinematic points, the light state contribution could vanish due to a delicate cancellation. This is no different from choosing a particular kinematic point in a scattering amplitude for which the coupling constants can be tuned so that the perturbative contributions vanish and only non-perturbative effects are non-zero. In particular, in CFTs with a sparse light spectrum, the light state contribution will generically not decrease in magnitude due to cancellations of neighboring phases as much as the heavy state contribution will.

We did not explore light-state dominance for the partition function with complex $\beta_L, \beta_R$, but studying this would be interesting. The spectral form factor $|Z(\beta + it)|^2$ provides information about information loss \cite{DyerG16}. The late-time behavior of the Virasoro block can be conveniently studied using the elliptic nome $q$, which has a natural description in terms of the pillow geometry, the $\mathcal{Z}_2$ quotient of the torus.

\section{Light-state dominance for derivatives}
\label{sec:derivatives}

In this section, we extend light-state dominance to derivatives of the correlator.  This is motivated by the fact that derivatives of correlators at the crossing symmetric point play a key role in the numerical bootstrap.  We begin by deriving  bounds on  derivatives of the partition function, $\partial^{2n} Z(\beta) $,  and then proceed to do the same for derivatives of the correlator, $\partial^{2n} G(z,\bar{z})$. For simplicity, we will work with $\beta_L = \beta_R$ and $z = \bar{z}$, but the derivations in previous sections will go through for the $\beta_L \neq \beta_R$ and $z \neq \bar{z}$ as before.

We find that the partition function and correlator admit a landscape of possible derivative bounds of varying strengths that can be derived using the procedure in this section. Ideal bounds are those for which the minimum value of $\Delta_c$ does not grow quickly in the number of derivatives and, unlike the bounds we will derive, are equally effective for operators of any dimension. It would be interesting to look for a convenient change of variables $z = f(y)$ in which derivative bounds are simpler to derive, and to consider applications to the numerical bootstrap and modular bootstrap.

\subsection{Revisiting Z: $(\partial_{\beta'})^{2n}$ }

We now establish light-state dominance for derivatives of the $2d$ partition function. We will derive bounds for even numbers of derivatives, as odd numbers of derivatives do not retain the positivity our arguments use. Consider for example
\begin{equation}
\partial_\beta Z_H' =  4 \pi^2 \sum_{E> \epsilon} \frac{ E}{\beta^2} e^{-\beta' E},
~~~~~~~
 \partial_\beta Z_H =  \sum_{E> \epsilon} - E e^{-\beta E}.
\end{equation}
These quantities have opposite signs.

Without loss of generality, suppose $ \beta > 2\pi$. We will consider the derivative operator $(\partial_{\beta'})^2$, which leads to derivative bounds in a particularly simple way in comparison to other derivative operators. Acting on the heavy contribution, the operator gives
\begin{equation}
(\partial_{\beta'})^2 Z_H =  \frac{1}{16 \pi^4}\sum_{E> \epsilon}  \beta^4 E(E-2/\beta ) e^{-\beta E}.
\end{equation}
We must take $\epsilon > 2 / \beta$ so that $E(E-2/\beta ) > 0$. Positivity for the heavy contribution is necessary to derive HKS-type bounds, but not for the light contribution. We then have
\begin{equation}
(\partial_{\beta'})^2 Z_H  \leq  \sum_{E> \epsilon}  \beta^4 E^2 e^{-\beta E}
\leq \beta^{4} e^{(-\beta+\beta') \epsilon}  \partial_{\beta'}^2 Z_H'.
\end{equation}
The function $\beta^4 e^{(-\beta+\beta') \epsilon} < 1$ for $\epsilon > \frac{4 \beta \ln \beta}{\beta^2-4\pi^2}$, which is larger than $2 / \beta$ for $\beta > 2\pi$, so we will take $\epsilon > \frac{4 \beta \ln \beta}{\beta^2-4\pi^2}$. It is important to note that this lower bound approaches 0 as $\beta \rightarrow \infty$. With $R \equiv \beta^4 e^{(-\beta+\beta') \epsilon} $, we have
\begin{equation}
(\partial_{\beta'})^2 Z_H \leq R (\partial_{\beta'})^2 Z_H'.
\end{equation}
and the next few steps are familiar. As $(\partial_{\beta'})^2 Z_{L}' \geq 0, $ we arrive at
\begin{equation}
\frac{R}{1-R}(\partial_{\beta'})^2 Z_L \geq  (\partial_{\beta'})^2 Z_H.
\label{ZDerivativeBound}
\end{equation}
Recall that the contribution of states with $E \in (0,2/\beta)$ to $(\partial_{\beta'})^2 Z_{L}$ is negative. \eqref{ZDerivativeBound} therefore places a bound on the number of states in this energy region, as too many states would turn the LHS negative, violating the positivity of the RHS. For large $\epsilon$, this leads to a bound on ``medium states'' $E \in (0,2/\beta)$ in terms of heavy and light data. For completeness, we state the bound:
\begin{equation}
\sum_{\substack{-c/12 \leq E \leq 0\\ 2/\beta \leq E \leq \epsilon} } E(E-2/\beta) e^{-\beta E} > \sum_{ 0< E <2 / \beta }  -E(E-2/\beta) e^{-\beta E}.
\end{equation}
The final result \eqref{ZDerivativeBound} provides a bound on derivatives of the heavy contribution in terms of derivatives of the light contribution.

We can now provide some intuition behind why we were able to find derivative bounds by modifying the original partition function argument in \cite{HartmanKS14}. The original argument can be understood in terms of competition between essential singularities at $\beta = 0, \infty$, where the precise balance is determined by $\beta$. Derivatives do not significantly change the singularity structure of $Z$, so for some $\beta$, the light and heavy state singularities must exchange dominance, and the same statement applies to their derivatives.

Not all squared derivative operators are positive when acting on positive functions, so it seems non-trivial that $(\partial_{\beta'})^2$ works. Positivity arises because both $e^{-\beta E},e^{-\beta' E}$ are concave up as functions of $\beta'$ for energies $E > 2/\beta$. The same property holds for $(\partial_{\beta'})^{2n}$, as $(\partial_{\beta'})^{2n} e^{-\beta E} = (a_0 \beta^{4n} E^{2n}+ a_1 \beta^{4n-1}E^{2n-1}+\ldots+ a_{2n+1}\beta^{2n+1}E) e^{-\beta E}$ with $a_0 > 0$, which is positive for large enough $E$. The $\beta < 2\pi $ case follows similarly. Note that for $\beta<2\pi$, one must use derivatives $(\partial_\beta)^{2n}$, as expected from modular invariance. This establishes a form of light-state dominance for $\partial^{2n} Z(\beta)$. 

It is now apparent that for large enough $\epsilon$, many derivative operators admit a heavy-state bound as long as they are positive acting on $E >\epsilon$ states. Generically these operators will be built from even derivatives $\partial^{2n}$ to achieve positivity. Acting with $\beta$-derivatives and multiplying by polynomials of $\beta, \beta'$ a finite number of times will not change the fact that for $\epsilon$ large enough and $\beta > 2\pi$, the suppression from $e^{-\beta E}$ relative to $e^{-\beta' E}$ will lead to an inequality $Z_H \leq R Z_H'$ with $R < 1$, and the rest of the proof will proceed as we have shown. In particular, the operator $D_\beta \equiv \beta \partial_\beta$ used in the modular bootstrap has light-state dominance for derivatives $D_{\beta}^{2n}$. Derivatives with respect to other variables can be analysed similarly. One challenge lies in finding derivative operators for which $\epsilon$ can be taken as small as possible.

\subsection{Light-state dominance for $(\partial_{z})^{2n}$}

We may follow the procedure that led to \eqref{ZDerivativeBound} to derive light-state dominance for $2n$-derivatives of the correlator. We will assume $z < 1/2$. With $\tilde{\Delta}' = \Delta'-2\Delta$,
\begin{equation}
(\partial_z)^2 \sum_{\Delta'} a_{\Delta'} z^{\tilde{\Delta}'} =  \sum_{\Delta'} a_{\Delta'} \tilde{\Delta}'(\tilde{\Delta}'-1) z^{\tilde{\Delta}'-2}
\leq
\left(\frac{z}{1-z}\right)^{\Delta_c-2}  (\partial_z)^2 \sum_{\Delta'} a_{\Delta'} (1-z)^{\tilde{\Delta}'},
\label{CorrelatorDerivatives}
\end{equation}
where $\Delta_c$ must be chosen such that $\tilde{\Delta}' (\tilde{\Delta}'-1) > 0$ and $\Delta_c \geq 2 $ for all heavy states. In particular, $\Delta_c > 2\Delta+2$ satisfies both conditions. The next steps are by now familiar and we arrive at the bound
\begin{equation}
\frac{R}{1-R}(\partial_{z})^2 (u^{-\Delta} G^s_L - v^{-\Delta} G^t_L) \geq  (\partial_{z})^2 (u^{-\Delta }G^s_H).
\end{equation}
Unlike in the zero-derivative case, $(\partial_{z})^2 ( v^{-\Delta} G^t_L) $ is not a sum of positive terms. States with $2\Delta > \Delta' > 2\Delta+1$ give negative contributions. In the special case that there are no operators in this region, we have $\frac{R}{1-R}(\partial_{z})^2 (u^{-\Delta} G^s_H) \geq  (\partial_{z})^2 (u^{-\Delta }G^s)$ and light-state dominance takes the same form as for the zero-derivative case, but we are interested in the general case. Defining $R_{n} \equiv \left(\frac{z}{1-z}\right)^{\Delta_c^{(n)}-2\Delta-2n}$, we can state the results for all even derivatives at once.
\begin{align*}
(\partial_z)^{2n} (u^{-\Delta} G_L(z))
&\leq (\partial_z)^{2n} (u^{-\Delta} G(z))
\\
&\leq  (\partial_z)^{2n} (u^{-\Delta} G_L(z))
+ \frac{R_n}{1-R_n} (\partial_{z})^{2n} (u^{-\Delta} G_L(z) - v^{-\Delta} G_L(1-z))
,
\numberthis
\label{DerivativeBoundSmallZ}
\end{align*}
where the cutoff $\Delta_c^{(n)} > 2\Delta+2n$. For $z>1/2$,
\begin{align*}
(\partial_z)^{2n} (v^{-\Delta} G_L(1-z) )
&\leq (\partial_z)^{2n} (u^{-\Delta} G(z))
\\
&
\leq (\partial_z)^{2n} (v^{-\Delta} G_L(1-z))
+
\frac{R_n^{-1}}{1-R_n^{-1}} (\partial_{z})^{2n} (v^{-\Delta} G_L(1-z) - u^{-\Delta} G_L(z))
.
\label{DerivativeBoundLargeZ}
\numberthis
\end{align*}
Higher derivatives probe heavier states. Notice that, unlike our bounds on $G(z, \bar{z})$ and $\partial^{2n} Z(\beta)$, in which case $c/12$ is the analogue of $2\Delta$, the derivative bounds are sensitive to the dimension of the external operator. For example, the $n=1$ bound requires $\Delta_c > 2$ even for $\Delta \ll 1$. The origin of this difference can be understood in terms of how the derivatives act on the singularities, or equivalently, on the powers of $z, e^{-\beta}$. $Z(\beta)$ has essential singularities in $\beta$, so $\beta$-derivatives do not change the strength of the singularity drastically. On the other hand, $G(z)$ has only power-law singularities in $z$, and so $z$-derivatives are in some sense more powerful, as they change the divergence of each term and the OPE singularity more dramatically.

As in the partition function case, similar bounds can be derived for a host of other derivative operators for sufficiently high $\Delta_c$. It would be useful to derive a bound that is valid for arbitrary $\Delta$ so that it may be applicable to the numerical bootstrap for generic operators.

\subsection{Light-state dominance and the numerical bootstrap}

In this section we will show that the heavy-state derivative bounds are meaningful at $z =\bar{z} = 1/2$, motivating possible future application to the numerical bootstrap. We will begin by revisiting the partition function and the modular bootstrap. In expanding $Z(\beta)-Z(\beta')$ about the self-dual point $\beta = 2\pi$ one obtains constraints on operator dimensions \cite{Hellerman09,FriedanK13}. The HKS light-state dominance results use the ratio of Boltzmann factors $R = {e^{(\beta'-\beta)\epsilon}}$, which is $1$ at $\beta  = 2\pi$, rendering the bound weaker and weaker as $\beta \rightarrow 2\pi$. However, the stronger upper bound is, for $\beta > 2\pi$,
\begin{equation}
Z_H \leq \frac{R}{1-R}(Z_L-Z_L').
\label{ModifiedHeavyStateHKSBound}
\end{equation}
Unlike the upper bound $\frac{R}{1-R} Z_L$, the upper bound \eqref{ModifiedHeavyStateHKSBound} is in general finite at $\beta = 2\pi$, as
\begin{equation}
\lim_{\beta \rightarrow 2\pi} \frac{R}{1-R}(Z_L-Z_L')
=
\lim_{\beta \rightarrow 2\pi} - \frac{ (Z_L-Z_L') \partial_{\beta}R + R \partial_\beta(Z_L-Z_L')}{\partial_\beta R}.
\end{equation}
The first term is zero and both $R/\partial_\beta R, \partial_\beta(Z_L-Z_L')$ are finite at $\beta = 2\pi$ for any density of states. The upper bound will also generically decay as we take $\epsilon$ large at fixed $c$. First, suppose that the only light state is the vacuum, $Z_L = e^{\frac{\beta c}{12}}$. The upper bound at $\beta = 2\pi$ is $Z_H \leq \frac{c}{12} \frac{e^{c/6}}{\epsilon}$ and including the additional states gives a similar result. The error incurred by neglecting heavy states can therefore be made arbitrarily small at $\beta = 2\pi$. However, $(\partial_{\beta'})^{2n} (Z_L-Z_L')$ is not zero at $\beta = 2\pi$, and so the upper bounds for derivative operators $(\partial_{\beta'})^{2n}$ are infinite at $\beta = 2\pi$.  It may be possible that other derivative operators will lead to a finite upper bound at $\beta = 2\pi$, and if so this bound may be relevant for the modular bootstrap, but we will not investigate this here.

We can now apply the same analysis to the bounds on the heavy-state contributions to the correlator. We immediately see that $(\partial_{z})^{2n} (u^{-\Delta} G^s_L - v^{-\Delta} G^t_L)|_{z=1/2} = 0$ for all $n$, and the previous analysis shows that the derivative bounds for correlator are finite at $z=1/2$.

We can further estimate the $\Delta_c$-dependence of the heavy state upper bound by computing the bound for the mean field theory correlator
\begin{equation}
G(z) = 1+z^{2\Delta}+\left(\frac{z}{1-z}\right)^{2\Delta}.
\end{equation}
In the s-channel, the $z^{2\Delta}$ contribution comes from the contribution of the $:\mathcal{O} \mathcal{O}:$ operator with dimension $2\Delta$. The remaining operators contribute to subleading terms in the $(1-z)^{-2\Delta}$ expansion. The general case $\Delta_c = 2\Delta+k+\epsilon$ with $0<\epsilon<1$ is
\begin{equation}
G^s_L = 1+z^{2\Delta}+z^{2\Delta} M_k,
~~~~~~~~~~~~
G^s_H = \left(\frac{z}{1-z}\right)^{2\Delta}-z^{2\Delta} M_k,
\end{equation}
where $M_k = \sum_{m=0}^{k} \frac{\Gamma(2\Delta+m)}{\Gamma(2\Delta)\Gamma(m+1)} z^m$  is the contribution of the non-identity states below the cutoff.
 
We will first examine the contribution of the identity operator. The upper bound on the heavy-state contribution to the correlator is
\begin{equation}
v^{\Delta} G_H(z) \leq \frac{R}{1-R}(v^\Delta G_L(z)-u^\Delta G_L(1-z)),
\end{equation}
with $R = \left(\frac{z}{1-z}\right)^{\Delta_c-2\Delta}$. When the identity is the only light operator exchanged, the upper bound on $G_H(z)$ approaches $ \frac{2 \Delta }{\Delta_c-2\Delta}$ as $z\rightarrow 1/2$. For comparison, this upper bound on the heavy-state contribution is equal to the upper bound obtained in \cite{LinSSWY15} for the whole correlator assuming only the identity was exchanged below $2\Delta$, which occurs in the $2d$ and $3d$ Ising models. Note that this upper bound decays in $\Delta_c$. The same behavior arises when we use the $z>1/2$ bound,
\begin{equation}
v^\Delta G_H(z) \leq \frac{1}{1-((1-z)/z)^{\Delta_c-2\Delta}} (u^\Delta G_L(1-z)-v^\Delta G_L(z)).
\end{equation}
We see that our upper bounds at $z=1/2$ decay as $1/\Delta_c$ or slower.

Now we consider the derivative bounds. For $z<1/2$,
\begin{equation}
(\partial_z)^{2n} (u^{-\Delta} G_H(z))
\leq  \frac{R_n}{1-R_n} (\partial_{z})^{2n} (u^{-\Delta} G_L(z) - v^{-\Delta} G_L(1-z)).
\end{equation}
The contribution of the identity gives an upper bound at $z = 1/2$ of $\frac{2^{3n+2\Delta+1}}{\Delta_c-2\Delta} \frac{(\Delta+n)!(2\Delta+2n-1)!!}{(\Delta-1)!(2\Delta-1)!!}$, which is suppressed in $\Delta_c$ and grows in $n$. The derivative bounds for $z>1/2$ approach the same value at $z=1/2$.

We can now apply the same analysis to the full mean field theory correlator. One can check explicitly by choosing various values of $k,n$ that the upper bound on the heavy-state contribution and its derivatives at $z=1/2$ decay in $\Delta_c$. To summarize, the upper bounds on derivatives of the heavy-state contribution at $z=1/2$ are finite for general CFTs, and we have checked only for the MFT correlator that the value of that upper bound decays in $\Delta_c$.

The derivative bounds may provide a method of quantifying the error in the numerical bootstrap procedure initiated by Gliozzi \cite{Gliozzi13,GliozziR14,GliozziLMR15}. The standard numerical bootstrap approach gives bounds the light-state spectrum (and OPE data) without needing to know the heavy-state data, which is generically unknown. This approach uses real OPE coefficients and so applies only to unitary theories. The Gliozzi approach relaxes this assumption. In this approach, one considers the theory truncated to a set of low-dimension primaries specified by $N$ independent parameters. Acting with linear functionals involving derivatives at the crossing-symmetric point $z = 1/2$ will produce $M$ equations. For $M = N$, we can solve for the $N$ parameters. Now we can check if this solution is stable, that is, unchanged when we increase $N$ or change the linear functionals used. A challenge in using this approach is that there is no built-in mechanism to validate that a stable solution indeed solves crossing to within a certain accuracy. Once a stable light-state solution is found, our bounds constrain the putative heavy-state contribution to the correlator and its derivatives. Further study of these bounds may lead to an algorithm for bounding the error on the $N$ parameters, but we leave this line of investigation to future work.

\section{Application to gapped and effective CFTs}
\label{sec:applications}

In a generic CFT we obtain a heavy state bound for any choice of cutoff $\Delta_c$, but $\Delta_c$ has no particular preferred value.   However, in some cases the existence of a gap in the spectrum of the theory does lead to a natural choice for $\Delta_c$.  We can then discuss the structure of the ``approximate CFT" in which contributions above the gap are discarded (working in the appropriate OPE channel), as well as its accuracy in reproducing correlators of the exact underlying theory that it approximates.   The most well known context for this discussion is that of large $N$ theories, in particular those arising in the AdS/CFT correspondence.  Large $N$ implies that the low lying operator spectrum (i.e. operators whose dimensions remain finite as $N\rightarrow  \infty$) can be separated into single and multi-trace sectors.  Connected $n\geq 3$ point correlation functions of the single trace operators are suppressed by powers of $1/N$.   CFTs with a local holographic dual have the further property that there is a twist gap below which all single trace operators have spin less than or equal to two.   More precisely, it was conjectured in \cite{HeemskerkPPS09} that the existence of such a gap implies the existence of an effective action in the bulk in which higher derivative operators are suppressed by powers of $\Delta_{gap}$.  In situations in which there is a gap in either the spectrum of all primary operators, or those of a particular type, such as those with spin less than or equal to two, it is natural to set $\Delta_c=\Delta_{gap}$, and in this section we will explore some implications of this.

\subsection{Effective field theories and CFTs}

It will be useful to recall some general facts about effective field theories.  The most familiar context arises in QFT in flat space.   Consider a theory with fields $\phi_L$ and $\phi_H$ with $m^2_L< m^2_H$.   In the kinematic regime $s=(p_1+p_2)^2<m_H^2$ we can expand a diagram representing tree level exchange of $\phi_H$ using
\bea\label{za}
{1\over s+m_H^2} = {1\over m_H^2} \sum_{n=0}^\infty \left( {-s\over m_H^2}\right)^n~,
\eea
which converges for $|s|<m_H^2$.   The term going as $s^n$ is reproduced in the effective theory by a quartic interaction with $2n$ derivatives, hence of schematic form $\p^{2n} \phi_L^4$.  Since the expansion (\ref{za}) is convergent, including more and more such operators in the effective Lagrangian yields arbitrary accuracy provided $s< m_H^2$.

The purpose of this elementary discussion is to contrast the situation with the analogous position space correlator $\langle \phi_L(x_1) \ldots \phi_L(x_4)\rangle$.  The point is that this correlator is not analytic in $1/m_H^2$ and so does not admit a convergent expansion in $m_H^2 x^2$; here $x$ denotes the overall length scale associated with the $x_i$.   This is evident even in the free field two-point function $\langle \phi_H(x) \phi_H(0)\rangle$, which behaves as $e^{-m_H|x|}$ at large distance.   As another example consider
\begin{equation}
G(x) = \int \frac{d^d p}{(2\pi)^d} \frac{e^{-i p x}}{(p^2)^2} \frac{1}{p^2+(m_H)^2},
\end{equation}
which corresponds to the correction to $\langle \phi_L(x) \phi_L(0)\rangle$ arising from a $\phi_L \phi_H$ vertex, and where we have set $m^2_L=0$.
Rewriting this as
\begin{equation}
G(x) = \frac{1}{(m_H)^2}\int \frac{d^d p}{(2\pi)^d} \frac{e^{-i p x}}{(p^2)^2}
-\frac{1}{(m_H)^4} \int \frac{d^d p}{(2\pi)^d} \frac{e^{-i p x}}{p^2}
+\frac{1}{(m_H)^4} \int \frac{d^d p}{(2\pi)^d} \frac{e^{-i p x}}{p^2+(m_H)^2},
\label{SpecificPositionSpaceEFT}
\end{equation}
we see that the first two terms belong to an expansion in $1/m_H^2$, while the last term, being proportional to the heavy field propagator, has an essential singularity at ${1\over m_H^2}=0$ on account of its $e^{-m_H |x|}$ dependence. If we tried to expand the last term in $1/m_H^2$ we would simply get a series of delta functions that vanish for $|x|\neq 0$.  The general expectation is, schematically,
\begin{equation}
G(x_i) = \sum_n a_n \left( \frac{1}{m_H^2 x^2}\right)^n + e^{-m_H |x|},
\label{GenericPositionSpaceEFT}
\end{equation}
where the $e^{-m_H |x|}$ term stands for any contribution with zero radius of convergence in the $1/m_H^2$ expansion.  As with any asymptotic expansion, including more and more terms in the $1/m_H^2$ expansion does not lead to arbitrarily good accuracy at fixed $m_H^2 x^2$; rather the error is necessarily of size $e^{-m_H |x|}$.   Conversely, keeping just the first $n$ terms in the expansion yields a result with error $O\big( m_H^2 x^2)^{-n}$ as $m_H^2 x^2 \rightarrow \infty$.

The preceding discussion also holds for position space correlators in AdS, which are dual to position space CFT correlators. Consider a Witten diagram for a light field involving the exchange of heavy fields.  As above, this correlator will admit an asymptotic expansion in $1/m_H^2$, and we can use this to obtain accurate results as $m_H^2 \rightarrow \infty$.   Furthermore, we can think of integrating out the heavy field to obtain a sum of contact interactions for the light field.   These contact interactions can reproduce the original correlator to an accuracy of at best $e^{-m_H |x|}$.

This can be made more explicit by thinking about the conformal block decomposition of a tree level diagram involving heavy field exchange.  Consider a theory with light fields $\phi_L$ and $\phi'_L$, a heavy field $\phi_H$, and cubic interactions $g \phi_L^2 \phi_H +g {\phi'}_L^2 \phi_H$.   The exchange diagram contributing to $\langle \phi_L \phi_L \phi'_L \phi'_L\rangle$ has a conformal block decomposition in terms of primaries of dimension $\Delta_n=2\Delta_{L} + 2n$, $\Delta'_n=2\Delta'_{L}+2n$ and $\Delta_{H} $, where as usual $m^2 =\Delta(\Delta-d)$.   These correspond to the  two-particle states built of the light particles, together with the heavy particle.  The heavy particle block, which is exponentially suppressed for large $m_H$, is precisely what is lost when one tries to integrate out the heavy field in favor of contact interactions for the light fields.

In more detail, the exchange diagram takes the form  \cite{GeodesicWittenDiagrams}
\bea
\mathcal{A}_{exchange} & =& \int_y \int_{y'} G_{b \partial}(x_1,y) G_{b \partial}(x_2,y)G_{bb}(y,y') G_{b \partial}(x_3,y') G_{b \partial}(x_4,y')
\cr &=& C_{LLH}C_{L'L'H} g_{\Delta_H,0}(z,\bar{z})+\sum_m P_m g_{\Delta_m,0}(z,\bar{z})+\sum_n P'_n g_{\Delta'_n,0}(z,\bar{z}),
\eea
where the coefficients of the double trace terms are
\begin{equation}\label{xa}
P_m = \left( \beta_{\Delta_m} \frac{a_m}{m^2_m-m^2_H} \right) \left( \beta_{\Delta'_m } \sum_n \frac{a'_m}{m^2_m-m'^2_n} \right)
\end{equation}
and similarly for $P'_n$.
The coefficients $a_{m}^{ij}, \beta_{\Delta ij}$ are defined in \cite{GeodesicWittenDiagrams} and will not be relevant for our purposes.

Now consider expanding in $1/m_H^2$.   $m_H$ appears in the coefficients $P_m$ and $P'_n$, as well as in the heavy block $g_{\Delta_H,0}(z,\zb)$.   The heavy block  decays exponentially for large mass.    This fact is easily understood by expressing the block as a geodesic Witten diagram \cite{GeodesicWittenDiagrams}, in which the dependence on $\Delta_H$ arises solely in the bulk-to-bulk propagator stretching between the two geodesics.  Since the bulk-to-bulk propagator behaves as $G(y,y')\sim e^{-\Delta_H \sigma(y,y')}$, where $\sigma $ is the geodesic distance, the block exhibits the falloff $g_{\Delta_H,0}(z,\zb)\sim e^{-\Delta_H \sigma_{min}(z,\zb)}$, where the function $\sigma_{min}(z,\zb)$ is identified with the minimal geodesic length connecting the two external geodesics.  Discarding the heavy exchange contribution leads to an error of size $e^{-\Delta_H \sigma_{min}(z,\zb)}$, consistent with our general results. 

Let us discard the heavy block and so define a truncated diagram
\bea\label{Atrunc}
{\cal A}_{trunc} = \sum_m P_m g_{\Delta_m,0}(z,\bar{z})+\sum_n P'_n g_{\Delta'_n,0}(z,\bar{z}) ~.
\eea
${\cal A}_{trunc}$ has an asymptotic series in $1/m_H^2$,  obtained simply by writing ${1\over m_m^2-m_H^2} = -{1\over m_H^2} \sum_k \left({m_m^2 \over m_H^2}\right)^k$ in (\ref{xa}).   The truncated diagram has a simple diagrammatic origin.   Since the bulk-to-bulk propagator is $G_{bb}(y,y') = \langle y| {1\over \nabla^2-m_H^2}|y'\rangle$, if we expand in $1/m_H^2$ we get the asymptotic expansion
\begin{equation}
\mathcal{A}_{exchange} \sim -{1\over m_H^2}  \sum_k  \int_y
 G_{b \partial}(x_1,y) G_{b \partial}(x_2,y) \left(\frac{\nabla_y^2}{m^2_H}\right)^k G_{b \partial}(x_3,y) G_{b \partial}(x_4,y).
\end{equation}
Term by term, this expansion agrees with the $1/m_H^2$ expansion of ${\cal A}_{trunc}$.

To establish this we write the expansion of the $P_m$ as
\bea\label{Ptil}
P_m =  -{1\over m_H^2} \sum_k \left({m_m^2 \over m_H^2}\right)^k \tilde{P}_m~.
\eea
The coefficients $\tilde{P}_m$ are in fact just the expansion coefficients of the basic contact diagram into double trace blocks,
\bea\label{Acon}
{\cal A}_{contact}&=& \int_y G_{b \partial}(x_1,y) G_{b \partial}(x_2,y) G_{b \partial}(x_3,y) G_{b \partial}(x_4,y)\cr &=& \sum_m \tilde{P}_m g_{\Delta_m,0}(z,\zb)    + \sum_n \tilde{P}'_n g_{\Delta'_m,0}(z,\zb).
\eea
Next we recall a basic identity used in the derivation of geodesic Witten diagrams,
\begin{equation}
G_{b \partial}(x_1,y) G_{b \partial}(x_2,y) = \sum_m a_m \phi_m (y),
\end{equation}
where $\phi_m (y)$ obeys
\begin{equation}
(\nabla^2 - m_m^2) \phi_m = 0
\end{equation}
except for a source localized on the geodesic running between $x_1$ and $x_2$.   Using this, each factor of $m_m^2$ appearing in (\ref{Ptil}) can be replaced by a $\nabla_y^2$, so that (\ref{Atrunc}) and (\ref{Acon}) together imply
\bea
{\cal A}_{trunc} \sim  -{1\over m_H^2}  \sum_k  \int_y
 G_{b \partial}(x_1,y) G_{b \partial}(x_2,y) \left(\frac{\nabla_y^2}{m^2_H}\right)^k G_{b \partial}(x_3,y) G_{b \partial}(x_4,y),
 \eea
 as claimed.

A related story pertains to the effect of heavy bulk field exchanges on the anomalous dimensions of light double trace operators, as was studied in \cite{FitzpatrickKPS10}.   A heavy t-channel exchange contribution to $\langle \phi_L(x_1) \phi_L(x_2) \phi_L(x_3) \phi_L(x_4)\rangle$, when expanded in terms of s-channel blocks, leads to anomalous dimensions $\gamma_{n,l}$ for double trace operators $[O_L O_L]_{n,l} \sim O_L (\p^2)^n \p_{\mu_1}\ldots \p_{\mu_l}O_L$.  We focus here on the spin-0 anomalous dimensions.  $\gamma_{n,0}$ has an expansion in $1/m_H^2$ with a finite radius of convergence.  In AdS$_{d+1}$, the expansion in the regime $n\gg \Delta_L$  takes the form $\gamma_{n,0} = \sum_k b_k \left(n\over m_H\right)^{d-3+2k}$.   This expansion is reproduced by  the sum of contact interactions arising by integrating out the massive field at tree level, with the $k$th term associated with a contact interaction with $2k$ derivatives.   Combining this with our previous discussion, we see that the truncated Witten diagram ${\cal A}_{trunc}$ will reproduce these terms.    The expansion has a finite radius of convergence, and breaks down when $n\sim m_H$.  The full expression for $\gamma_{n,0}$ displays a resonance type behavior near $n\sim m_H$,  and then eventually decays as $\gamma_{n,0} \sim 1/n^{5-d}$, assuming $d<5$.  In summary, the situation for anomalous dimensions is similar to that of correlation functions in momentum space, in the sense that heavy particles yield a low energy (or low $n$) expansion with a finite radius of convergence.  We expect the same to hold for double trace OPE coefficients.

We now consider the relation between the truncated correlator ${\cal A}_{trunc}$ and the light state correlator $G_L$ we defined previously.  They are not same, since $G_L$ is obtained by discarding exchanged operators above a cutoff, while ${\cal A}_{trunc}$ retains the contribution of double trace operators of arbitrarily high dimension.  However, we should recall that ${\cal A}_{trunc}$ is defined as an asymptotic expansion, and like all such expansions we should retain only a finite number of terms to obtain the best accuracy at fixed expansion parameter.  In the present case, the usual rules tell us  that we should retain only those double trace operators whose dimension is less than the heavy exchanged operator.   Upon doing so, ${\cal A}_{trunc}$ agrees with $G_L$.

\subsection{Inverting the gap}

As reviewed above, integrating out a heavy particle yields contact interactions for the light particles with coefficients suppressed by the heavy mass.   On the most general grounds, the total contribution of heavy particles above a gap to the contact interactions could be unsuppressed, either due to the multiplicity of such particles or due to their large coupling constants with light particles.  One would like to establish under what conditions this can be ruled out, so that the  virtual heavy particle contributions are indeed suppressed at distance scales $l> \Delta_{gap}^{-1}$.

To streamline the discussion we will consider the following scenario, which is meant to correspond to a string background AdS$\times M$, where $l_{AdS} \gg \sqrt{\alpha'}$ but $l_M \sim \sqrt{\alpha'}$, so that $\Delta_{gap} \sim l_{AdS}/\sqrt{\alpha'}$.  We henceforth set $l_{AdS}=1$.   We assume there are a finite number of single trace primaries below the gap, all of spin no greater than $2$.  We consider varying $\Delta_{gap}$, and assume that  the properties (i.e. dimensions, spins, and OPE coefficients)  of the light single trace spectrum remain unchanged as we do so.   We are interested in characterizing the dependence on $\Delta_{gap}$ of the OPE coefficients  $C_{OO' [OO']_{n,l}}$, which are related to the coefficients of bulk contact interactions.

As discussed in \cite{OPEInversion}, at least at order $1/N$ (corresponding to tree level in the bulk) this problem can be efficiently addressed using the Lorentzian inversion formula, which is used to convert t-channel exchanges into s-channel OPE data. One reason the inversion formula is useful here is that it involves the double discontinuity (dDisc) of the correlator, and at order $1/N$ the dDisc only receives contributions from single trace primary exchanges. So, even though the Witten diagram for heavy particle exchange corresponds to the exchange of both light and heavy primaries, only the heavy part matters.  We can therefore use bounds on heavy state contributions to correlators to place bounds on $C_{ OO' [OO']_{n,l}}$ OPE coefficients, and hence on the contact interactions.   The other key fact is that dDisc is positive and bounded by the Euclidean correlator, dDisc$(G) \leq G_E$.

For brevity, we will assume familiarity with OPE inversion formula computations, for example \cite{OPEInversion,AldayHL2017,AldayC17,LiuPRS18,KrausSS18}. For a pedagogical introduction to such calculations, see \cite{KrausSS18}.   We will also not pay close attention to overall factors that are independent of $\Delta_{gap}$.

On the principal series, the inversion formula is
\begin{equation}\label{zw}
c(\Delta,J) = \frac{\kappa_{\Delta+J}}{4} \int dz d\bar{z} \mu(z,\bar{z}) g_{H,\bar{H}}(z,\bar{z}) \dDisc(G(z,\bar{z})),
\end{equation}
where $H = 1-h$ with $h = \frac{\Delta-J}{2}$ and $\bar{H} = \bar{h} = h+J$. The block $g_{H, \bar{H}}(z, \bar{z}) \sim z^H \bar{z}^{\bar{H}}$ for small $z, \bar{z}$ and $\log(1-z)\log(1-\bar{z})$ for $ (z,\zb) \sim (1,1)$. The integral is not convergent for negative $H$ even though this corresponds to positive $h$, the physical regime. Following \cite{OPEInversion}, we will bound the contribution of heavy states to $c(\Delta,J)$ in the region in which the integral converges.

As the t-channel cutoff $\Delta_{gap} \rt \infty$,
an asymptotic estimate for the Euclidean correlator $G_E$ is obtained from tauberian theorems \cite{PappadopuloRER12} by demanding that the tail reproduce the leading s-channel OPE singularity.  The analysis in \cite{OPEInversion} then leads to a bound dDisc$(G)|_{\rm heavy} \leq e^{-\Delta_{gap}(\sqrt{z}+\sqrt{\zb})}$, and carrying out the inversion integral then leads to bounds of the form
\bea\label{zz}  |c(\Delta,J)_{\rm heavy}| \leq {1  \over (\Delta_{gap}^2)^{J-1} }~.
\eea
This formula has been obtained for unphysical operator dimensions $\Delta= {d\over 2}+i\nu$.   The physical $\Delta$ lie on the real axis, and show up as poles in $c(\Delta,J)$, with the residues of the poles yielding the corresponding OPE coefficients.  For present purposes, what is relevant is that contributions to light OPE coefficients are extracted from $1/N$ corrections to residues of poles on the real axis. We can extract the residue from a contour integral using Cauchy's theorem, but to apply (\ref{zz}) we need to assume that the bound on $|c(\Delta,J)_{\rm heavy}|$ can be extended to the integration contour, an assumption we will make henceforth \footnote{The behavior of $c(\Delta,J)$ in the complex $\Delta$ plane requires careful analysis and is not fully understood. Here is an alternate argument. In \cite{MukhametzhanovZ18}, it was shown that $c(\Delta,J)$ grows at most polynomially in $|\Delta|$ in the complex $\Delta$ plane, up to familiar ambiguities at finite $J$. This leads to a dispersion relation that relates $c(d/2+i\nu,J) + \text{``extra''}$ to the OPE coefficient density $\rho(\nu)$. The ``extra'' terms do not depend on the external operator dimension or data in the theory and correspond to $\rho(\nu)$ contributions which are zero when integrated against the blocks, so we can neglect them in analysing the $\Delta_{gap}$ dependence. It now follows that bounding $c(\Delta,J)$ on the principal series also bounds OPE coefficients.}.
 Under these assumptions, the bound on OPE coefficients translates to the statement that a contact interaction coupling to spin-$J$ double trace operators  receives a bounded contribution from heavy particle exchanges as $\Delta_{gap}\rt \infty$.

We will proceed somewhat differently, instead using our heavy state bound to bound the OPE coefficients at fixed $\Delta_{gap}$. Since the basic input is a bound on the correlator in terms of the light correlator $G_L$, in order for this to be useful we need a bound on $G_L$.  $|G_L|$ is of course unbounded on the Euclidean plane due to OPE singularities.  However, as discussed in section 3.1, since these are the only singularities, we can write
\bea\label{zy}  |G_L(z,\zb)| \leq A |G_{MFT}(z,\zb)|
\eea
for some constant $A$, where $G_{MFT}(z,\zb)$ is the mean field theory correlator consisting of products of two-point functions.  The optimal value of $A$ is of course theory specific, and the larger $A$ is the larger $\Delta_{gap}$ will need to be in order that the heavy state contribution is small.

We return to  (\ref{zw}) and apply our bounds.   We define
\bea R_z = \frac{z}{1-z}.
 \eea
For large $\Delta_c$, the heavy state bounds in \eqref{LightStateBoundZZbarEqual}, \eqref{LowTempZZbarDifferent}, and \eqref{HighTempZZbarDifferent} take essentially the same form. To summarize the leading contributions,
\begin{align*}
z,\bar{z} < 1/2: & \quad \quad G(z, \bar{z}) \leq G_L(z, \bar{z})\left(1+(R_z R_{\bar{z}})^{\Delta_c}\right)
\\
z<1-\bar{z}< 1/2: & \quad \quad G(z, \bar{z}) \leq G_L(z, \bar{z})\left(1+R_z^{h_c} ( R_{\bar{z}}/R_z)^{\Delta_O} \right)
\\
1-\bar{z}<z< 1/2: & \quad \quad G(z, \bar{z}) \leq G_L(1-z,1-\bar{z})(R_z R_{\bar{z}})^{\Delta_O}\left( 1+ R_{\bar{z}}^{-h_c} ( R_{\bar{z}}/R_z)^{\Delta_O}  \right)
\\
z,\bar{z} > 1/2: & \quad \quad G(z, \bar{z}) \leq (R_z R_{\bar{z}})^{\Delta_O} G_L(1-z, 1-\bar{z})\left(1+(R_z R_{\bar{z}})^{-\Delta_c}\right)
.
\numberthis
\end{align*}
We have used $\Delta_O$ to denote the external operator dimension to avoid confusion. The $z, 1-\bar{z}>1/2$ bounds will be unnecessary due to the $z \leftrightarrow \bar{z}$ symmetry of the inversion formula integrand.

We now bound $G_L$ by $AG_{MFT}$.   The latter contains three terms, corresponding to the three disconnected contributions.  We now focus on the simplest term, $G_L =1$, which suffices to illustrate the point.  Alternatively, we can consider a correlator of pairwise identical operators such that this is the only contribution.
Proceeding, the leading contribution that depends on $\Delta_c$ is, with $\Delta'=\Delta_c-2\Delta$,
\begin{align*}
c(\Delta,J) \leq & A\int dz d\bar{z} \mu(z, \bar{z}) g_{H,\bar{H}}(z,\bar{z})
\bigg(
(R_z R_{\bar{z}})^{\Delta'} \bigg|_{z,\bar{z} < 1/2}
+
R_z^{h_c} ( R_{\bar{z}}/R_z)^{\Delta_O}
\bigg|_{z<1-\bar{z}< 1/2}
\\
&+
 R_{\bar{z}}^{-h_c} R_{\bar{z}}^{2\Delta_O} \bigg|_{1-\bar{z}<z< 1/2}
+
(R_z R_{\bar{z}})^{\Delta_O} (R_z R_{\bar{z}})^{-\Delta'}\bigg|_{z,\bar{z} > 1/2}
\bigg)
.
\numberthis
\end{align*}
Here we have broken the integration domain into four regions as indicated by the $|_{\cdots}$ notation.

Consider the first term. The factor $(R_z R_{\bar{z}})^{\Delta_c}$ is sharply peaked around $z=\zb={1\over 2}$, falling off exponentially as we move away from this point, $(R_z R_{\bar{z}})^{\Delta_c} \sim e^{\Delta_c(z+\zb-1)}$. We can therefore evaluate the rest of the integrand at this point, except for a factor $|z-\zb|^{d-2}$ in $\mu(z,\zb)$ than vanishes there. Writing $\mu(z,\zb)= |z-\zb|^{d-2} \tilde{\mu}(z,\zb)$ we then have the estimate
\bea
\label{zu}
\int dz d\bar{z} \mu(z, \bar{z}) g_{H,\bar{H}}(z,\bar{z})
(R_z R_{\bar{z}})^{\Delta'} \bigg|_{z,\bar{z} < 1/2}  &\approx & C\int_0^{1/2}\int_0^{1/2} dzd\zb |z-\zb|^{d-2} e^{\Delta_c(z+\zb-1)}\cr
& \approx &{\# C\over (\Delta_c)^d}~.
\eea
with $C= \tilde{\mu}({1\over 2},{1\over 2}) g_{H,\overline{H}}({1\over 2},{1\over 2})$.   The other terms behave similarly.

To make things more explicit and to highlight some other important issues we specialize to $d=2$, for which we  have
\begin{align*}
c(\Delta,J) \leq & A\int dz d\bar{z} z^{H-2}\bar{z}^{\bar{H}-2} \F(H,H,2H,z)\F(\bar{H},\bar{H},2\bar{H},\bar{z}) + (H \leftrightarrow \bar{H})
\\
&
\times
\bigg(
(R_z R_{\bar{z}})^{\Delta'} \bigg|_{z,\bar{z} < 1/2}
+
R_z^{h_c} ( R_{\bar{z}}/R_z)^{\Delta_O}
\bigg|_{z<1-\bar{z}< 1/2}
\\
&+
 R_{\bar{z}}^{-h_c} R_{\bar{z}}^{2\Delta_O} \bigg|_{1-\bar{z}<z< 1/2}
+
(R_z R_{\bar{z}})^{\Delta_O} (R_z R_{\bar{z}})^{-\Delta'}\bigg|_{z,\bar{z} > 1/2}
\bigg)
.
\numberthis
\end{align*}
The integrals factorize in $z, \bar{z}$ for each term. The dependence of each term on $\Delta_c ~ h_c $ comes from $I_\pm (H),I_\pm(\bar{H})$ where
\begin{align*}
I_-(H)=&\int_0^{1/2} dz z^{H-2} \F(H,H,2H,z) \cdot z^{\Delta_c}(1-z)^{-\Delta_c}
\\
I_+(H)=&\int_{1/2}^{1} dz z^{H-2} \F(H,H,2H,z) \cdot z^{-\Delta_c}(1-z)^{\Delta_c}.
\numberthis
\end{align*}
These integrals fall off like $1/\Delta_c$.

To obtain $c(\Delta,J)$ for real $\Delta > 0$ and therefore compute OPE coefficients, we must continue $H, \bar{H}$ away from regions in which the integrals converges, but it is clear that the integrand will still decay in $\Delta_c$. For example, using the Cauchy-Schwarz inequality we can isolate the decay in $\Delta_c$ according to $\int_0^{1/2} dz z^{H-2} \F(H,H,2H,z) \cdot z^{\Delta_c}(1-z)^{-\Delta_c} \leq \sqrt{\int_0^{1/2} dz z^{H-2} \F(H,H,2H,z) \int_{0}^{1/2} dz z^{\Delta_c}(1-z)^{-\Delta_c}}$.  We then have $ \int_0^{1/2} dz z^{\Delta_c} (1-z)^{-\Delta_c} = 2\F(1,\Delta,2+\Delta,-1)/(1+\Delta),$ which decays to $0$ as $\Delta_c \rightarrow \infty$.

Taking $\Delta_c = \Delta_{gap}$, we conclude that $C_{\mathcal{O} \mathcal{O} \mathcal{O}_{\Delta,J}}^2|_{\rm heavy}  < A f(\Delta_{gap})$, where $f(\Delta_{gap})$ decays to zero as $\Delta_{gap}\rt \infty$.  It then follows that the heavy state contribution to light contact interactions similarly decays.  As we already noted, without knowing the optimal value of $A$, which depends on the theory, we cannot say when the large $\Delta_{gap}$ asymptotics set in.  Also, our bounds hold for finite $\Delta_{gap}$, but it is rather crude in that it doesn't provide us with the detailed dependence on $\Delta$ and $J$.

\subsection{Light-state dominance with non-identical operators}

We will now study features of light-state dominance with non-identical external operators by considering the correlator $\braket{\mathcal{O}_1(0)\mathcal{O}_1(z,\zb)\mathcal{O}_2(1)\mathcal{O}_2(\infty)}$ of scalar operators of dimension $\Delta_{1,2}$. Using our existing results, we can derive a bound on the heavy state contribution to this correlator. We will derive the bound using the conformal block decomposition, as we will use the bound to comment on the case of heavy-light correlators and contrast light-state dominance with $2d$ vacuum-block dominance.

We will work with $z = \bar{z}$ for simplicity and take $z<1/2$.  The conformal block expansion in the s-channel is
\begin{equation}
\braket{\mathcal{O}_1\mathcal{O}_1\mathcal{O}_2\mathcal{O}_2} = z^{-2\Delta_1}
\sum_{\Delta_p, l_p} C_{11 p} C_{22 p} g^{11 \rightarrow 22}_{\Delta_p,l_p}(z).
\label{MixedCorrelator}
\end{equation}
The blocks that appear are independent of $\Delta_{1,2}$, so we henceforth write $g^{11 \rightarrow 22}_{\Delta_p,l_p}(z)= g_{\Delta_p,l_p}(z)$.

To derive a heavy state bound, we first write
\begin{equation}
|\braket{\mathcal{O}_1\mathcal{O}_1\mathcal{O}_2\mathcal{O}_2}^s_H| \leq z^{-2\Delta_1}
\sum_{\Delta_p > \Delta_c} |C_{11 p} C_{22 p}| |g_{\Delta_p,l_p}(z)|.
\end{equation}
Since $(A\pm B)^2\geq 0$ we have
\begin{equation}
|C_{11 p}C_{22 p}| \leq \frac{1}{2} \left( (C_{11p})^2 + (C_{22p})^2\right)~,
\label{TriangleInequality}
\end{equation}
and so
\begin{equation}
|\braket{\mathcal{O}_1\mathcal{O}_1\mathcal{O}_2\mathcal{O}_2}^s_H| \leq
z^{-2\Delta_1}
\frac{1}{2}\sum_{\Delta_p > \Delta_c} \left( (C_{11 p})^2 + (C_{22 p})^2 \right) g_{\Delta_p,l_p}(z),
\end{equation}
where we used positivity of $ g_{\Delta_p,l_p}(z)$.   The two terms are precisely the heavy state contributions to $\braket{\mathcal{O}_1\mathcal{O}_1\mathcal{O}_1\mathcal{O}_1}$ and $\braket{\mathcal{O}_2\mathcal{O}_2\mathcal{O}_2\mathcal{O}_2}$, and so we deduce
\begin{equation}
|\braket{\mathcal{O}_1\mathcal{O}_1\mathcal{O}_2\mathcal{O}_2}^s_H|
\leq
{1\over 2}\braket{\mathcal{O}_1\mathcal{O}_1\mathcal{O}_1\mathcal{O}_1}^s_H
+
{1\over 2} z^{2(\Delta_2-\Delta_1)}\braket{\mathcal{O}_2\mathcal{O}_2\mathcal{O}_2\mathcal{O}_2}^s_H.
\end{equation}
We will take $\Delta_c > \text{max}( 2\Delta_1, 2\Delta_2)$. In this case, light-state dominance gives
\begin{equation}
|\braket{\mathcal{O}_1\mathcal{O}_1\mathcal{O}_2\mathcal{O}_2}^s_H|
\leq
\frac{R}{1-R}
\left(
\braket{\mathcal{O}_1\mathcal{O}_1\mathcal{O}_1\mathcal{O}_1}^s_L
+
z^{2(\Delta_2-\Delta_1)}\braket{\mathcal{O}_2\mathcal{O}_2\mathcal{O}_2\mathcal{O}_2}^s_L\right),
\end{equation}
where $R = \left(\frac{z}{1-z}\right)^{\Delta_c-\text{max}( 2\Delta_1, 2\Delta_2)}$.

The smallest possible value for $\Delta_c$ is set by the dimension of the heavier operator. Therefore, if we were to consider a heavy-light correlator, meaning that either of $\Delta_{1,2}$ is comparable to or larger than $c$, then we do not arrive at a useful bound.  On the other hand, in certain situations we do expect that such correlators are dominated by light states.  In particular, in the case of AdS$_3$/CFT$_2$ correlators can be dominated by the Virasoro vacuum block, corresponding to pure gravity in the bulk sourced by the heavy operator.  However, the validity of restricting to the Virasoro vacuum block depends on details of the theory (such as the absence of comparable contributions from conserved currents), and hence requires a separate analysis from that considered here.

\section{Discussion}
\label{sec:discussion}

Understanding the relative contributions of light and heavy states to CFT correlation functions is of interest both from a general CFT perspective as well as  for quantum gravity via AdS/CFT.
We have used crossing symmetry to derive light-state dominance in arbitrary CFT$_d$. Our results provide a general mechanism by which low-energy states furnish a good description of CFT observables. Light-state dominance also sheds light on the effect of a parametrically large gap. We have found agreement with effective field theory expectations in AdS/CFT and addressed the conjecture that a higher-spin gap leads to bulk locality \cite{HeemskerkPPS09}. We will conclude by mentioning some future directions.

\begin{itemize}
	\item \textbf{Higher-point correlators}
Our results were obtained for the four-point function, but we expect similar results for higher-point correlators.  Our main proof of light-state dominance followed from positivity of expansion coefficients, a crossing relation, and the fact that states entered as $z^{h_p} \bar{z}^{\bar{h}_p}$ and so had Boltzmann-type factors that were monotonically decreasing in $h_p, \bar{h}_p$.   The details change for higher-point correlators and so the procedure would need to be modified accordingly.

Studying $n$-point functions would in principle allow us to repeat our AdS/CFT analysis for $n$-point tree exchange diagrams. This would lead to constraints on towers of $n$-point higher-derivative contact interactions and we expect their coefficients to be similarly suppressed by powers of $\Delta_{gap}^{-1}$. While the machinery for carrying out these steps is not yet in place, we expect that in principle the story will be the same as for the four-point function.

	\item \textbf{Mixed correlators} The arguments we have used for light-state dominance relied heavily on the fact that the external operators were identical. In the special case of $\braket{\mathcal{O}_1 \mathcal{O}_1 \mathcal{O}_2 \mathcal{O}_2}$, we showed that the heaviest operator in the correlator determines the value of $\Delta_c$. We expect this to hold generically. It is possible that studying correlators of sums of operators, for example $\braket{(\mathcal{O}_1+ \mathcal{O}_2)^4}$ may allow one to extend light-state dominance to mixed correlators.

	\item \textbf{OPE spectral density:}
We have derived bounds on the total heavy-state contribution to the correlator, but one can use these bounds to investigate the OPE spectral density - squared OPE coefficients weighted by the density of states. By performing the inverse Laplace transform one can translate knowledge of the correlator into knowledge of the OPE spectral density, as carried out in detail in \cite{KrausM16,DasDP17,MukhametzhanovZ18}. One may also apply the approach of \cite{KimKO15}. As our heavy-state bounds in principle access lower dimensions than tauberian theorem methods, it would be interesting if our bounds could be used to clarify the validity of the OPE spectral density estimates obtained in \cite{MukhametzhanovZ18}, which surprisingly agree with known examples to much lower dimensions $\Delta_p$ than expected. As we have discussed, unless we specify the light data, the tauberian theorem methods and our heavy-state bounds are closely related. As in HKS, it would be interesting to explore the space of theories implied by certain light-state sparseness conditions.

	\item \textbf{$2\Delta$ and particle creation:} The quantity $2\Delta$ appears throughout our work as a lower bound on $\Delta_c$, and plays a similar role in the bounds derived in \cite{KimKO15,LinSSWY15}. It may be worth mentioning two contexts for this.  First, at large $N$ the dimension $2\Delta$ is special in that an operator of this dimension can mix with the double trace operator $:\mathcal{O}\mathcal{O}:$, and this fact shows up in bulk perturbation theory; e.g. \cite{FreedmanMMR98}.  Similarly, in flat space scattering the energy $2m$ marks the threshold for producing two-particle states. One might expect, for example, that studying light-state dominance in Mellin space and then taking the flat space limit would connect the $2\Delta$ threshold to that of particle creation in a concrete way.

\end{itemize}

\section*{Acknowledgements}

We wish to thank Tom Hartman, Tristan McKinney, Ben Michel, Eric Perlmutter, David Simmons-Duffin, and River Snively for enlightening discussions. We thank Julio Martinez for valuable discussion throughout the completion of this work and comments on a draft. We also wish to thank the participants of the Simons Collaboration on the Non-perturbative Bootstrap workshop held in Caltech in 2018 for useful, stimulating discussions. P.K. is supported in part by NSF grant PHY-1313986.

\bibliographystyle{ssg}
\bibliography{refs}

\end{document}